\pdfoutput=1 
\documentclass[12pt, draftclsnofoot, onecolumn]{IEEEtran}
\usepackage{subfigure}
\normalsize
\ifCLASSINFOpdf
\else
\fi

\usepackage{cite}
\usepackage{amsmath,amssymb,amsfonts}
\usepackage{algorithm}               
\usepackage{algorithmic}             
\usepackage{graphicx}
\usepackage{textcomp}
\usepackage{xcolor}
\usepackage{setspace}
\begin{document}
%
\title{Delay Analysis of Wireless Federated Learning Based on Saddle Point Approximation and Large Deviation Theory}
%
%
%

\author{\IEEEauthorblockN{Lintao Li, Longwei Yang, Xin Guo, \emph{Member, IEEE}, Yuanming Shi, \\ \emph{Senior Member, IEEE}, Haiming Wang, Wei Chen, \emph{Senior Member, IEEE}, \\and Khaled B. Letaief, \emph{Fellow, IEEE}}
\thanks{This work is supported in part by the National Natural Science Foundation of China under Grant No. 61971264, the Beijing Natural Science Foundation under Grant No. 4191001, and Lenovo Research. The preliminary work has been accepted by IEEE ICC 2021.} 
\thanks{Lintao Li, Longwei Yang and Wei Chen are with the Department of Electronic Engineering, Tsinghua University, Beijing 100084, China, and also with the Beijing National Reasearch Center for Information Science and Technology, Tsinghua University, Beijing 100084, China (email: llt20@mails.tsinghua.edu.cn; ylw18@mails.tsinghua.edu.cn; wchen@tsinghua.edu.cn).}
\thanks{Xin Guo and Haiming Wang are with Lenovo Research, Beijing 100094, China (email: guoxin9@lenovo.com; wanghm14@lenovo.com).}
\thanks{Yuanming Shi is with the School of Information Science and Technology, ShanghaiTech University, Shanghai 201210, China (email: shiym@shanghaitech.edu.cn).}
\thanks{Khaled B. Letaief is with the School of Engineering, Hong Kong University of Science and Technology, Clear Water Bay, Hong Kong, and also with Peng Cheng Laboratory, Shenzhen 518066, China (email: eekhaled@ust.hk).}\vspace{-4mm} }

\maketitle
\vspace{-17mm}
\begin{abstract}
\vspace{-3mm}

Federated learning (FL) is a collaborative machine learning paradigm, which enables deep learning model training over a large volume of decentralized data residing in mobile devices without accessing clients' private data. Driven by the ever increasing demand for model training of mobile applications or devices, a vast majority of FL tasks are implemented over wireless fading channels. Due to the time-varying nature of wireless channels, however, random delay occurs in both the uplink and downlink transmissions of FL. How to analyze the overall time consumption of a wireless FL task, or more specifically, a FL’s delay distribution, becomes a challenging but important open problem, especially for delay-sensitive model training. In this paper, we present a unified framework to calculate the approximate delay distributions of FL over arbitrary fading channels. Specifically, saddle point approximation, extreme value theory (EVT), and large deviation theory (LDT) are jointly exploited to find the approximate delay distribution along with its tail distribution, which characterizes the quality-of-service of a wireless FL system. Simulation results will demonstrate that our approximation method achieves a small approximation error, which vanishes with the increase of training accuracy.

\end{abstract}

\vspace{-6mm}
\begin{IEEEkeywords}

\vspace{-3mm}
Wireless federated learning, task-oriented communications, delay analysis, saddle point approximation, large deviation theory, extreme value theory.
\end{IEEEkeywords}

%
\IEEEpeerreviewmaketitle

\section{Introduction}
\vspace{-1mm}
%
%
%
%

The phenomenal development of wireless networking and machine learning technologies yields huge volume of sensory data at the wireless edge networks \cite{cm6}. Meanwhile, as the computational power and storage of mobile devices keep growing, mobile edge computing (MEC) \cite{1} is becoming a crucial technology for future networks to enable ultra-low power and ultra-low latency applications at the edge \cite{cm2}. By integrating federated learning (FL) and edge computing with enhanced privacy and security guarantees, federated edge learning has recently been receiving increasing attention to provide services and intelligence at the edge. This is achieved by training machine learning models across a fleet of participating distributed mobile devices without transferring their local private data to a remote centralized server at either the edge or cloud. FL thus becomes a key technique to support the paradigm shift from “connected things” to “connected intelligence” in 6G, where humans, things, and intelligence are intertwined within a hyper-connected cyber-physical world \cite{2,6Gwalid}. This inspires extremely exciting 6G applications, including industrial Internet of Things (IIoT), Internet of Vehicles (IoV), and healthcare \cite{cm1,cm3,cm5}. However, the deployment of FL in wireless networks, poses unique challenges in terms of system heterogeneity, statistical heterogeneity, and trustworthiness \cite{cm4,3}. In particular, due to the fading nature and limited resources of wireless channels, communication-efficiency becomes a key performance indicator to implement FL at scale in wireless networks with low-latency, privacy and security guarantees.

To address the communication challenges in wireless FL, a growing body of recent works has demonstrated the effectiveness of joint optimizing communication, computation and learning across wireless networks. The analysis and design of wireless FL systems can be typically divided into two categories: analog and digital communication protocols. Specifically, by exploiting the signal superposition property of a wireless multiple-access channel, analog FL has recently received particular interest to implement low-latency global model aggregation (e.g., the weighted average function computation) via over-the-air computation (AirComp). This scheme can significantly reduce the communication bandwidth consumption and improve spectrum efficiency via concurrent transmission of locally updated models.
However, the channel noise in the AirComp based model aggregation procedure yields a completely different type of distributed FL algorithms with iterative noise. It turns out to be difficult to characterize the convergence rates, complexity, optimality and statistical behaviors of the noisy FL algorithm iterates, for which various communication-efficient analog distributed algorithms were developed with convergence guarantees, e.g., analog gradient methods \cite{tsp1}, quantization schemes \cite{huanggunduz}, and gradient sparsification approaches \cite{mma1}. 
Moreover, resource allocation becomes critical to improve the learning convergence rate and model prediction accuracy in wireless FL, including the transmission power control \cite{mma1, shencong, jsacnew,xujie2}, model aggregation beamforming \cite{mma3, shiym}, and device scheduling \cite{shiym,zhu2019broadband,RIS,xia2020fast}. However, analog wireless FL system normally requires a strict synchronization at the symbol level and the aggregation procedure of the underlying distributed learning algorithms is inherently corrupted by channel noise.

In contrast, digital wireless FL systems are able to leverage advanced coding schemes and mobile edge computing techniques, thereby achieving channel noise robustness and low-latency transmission during the distributed wireless FL procedure. Various communication network architectures, e.g., single-server, hierarchical and decentralized wireless FL systems, have recently been proposed to improve the communication efficiency and learning performance. In particular, numerous research efforts have been made to minimize the training delay of single-server FL systems, including device scheduling policies for mitigating the stragglers (i.e., the training bottleneck caused by the slowest participating devices) \cite{ poor1,huangkb1,mma4,hjw1,aoi,Niu,Chenmz}, resource allocation for improving the transmission efficiency \cite{Niu,Chenmz,algorithm,bennis2}, adaptive aggregation for dynamically controlling the aggregation frequency \cite{JSAC2019} or calibrating aggregation weights \cite{poor2}, batchsize selection for improving learning efficiency \cite{yu}, and incentive mechanisms for reducing the side effects of information asymmetry \cite{incentive2,incentive1}. To further  address the heterogeneity in terms of computation and communication capabilities across devices, hierarchical FL was advocated to orchestrate nodes for cooperative learning \cite{dai}. To this end, a partial model aggregation approach was proposed in \cite{kl} to cope with communication bottlenecks, while a collaborative FL was proposed in \cite{cmz} for enabling on-device learning  with less reliance on a central server. Besides, to improve the system robustness and alleviate the straggler effect, a decentralized device-to-device communication enabled FL network architecture was proposed in \cite{xujie}, which is supported by radio resource allocation and computation loads adjustment.

Although there have already been a number of published works related to the delay minimization in both analog and digital transmission schemes, a delay distribution analysis has not yet been carried out for wireless FL systems. Indeed, delay distribution is critical and has profound implications for resource budgeting and timeout probability assessment under delay constraints. In this paper, we shall propose a novel framework to characterize the delay distribution analysis in wireless FL systems, for which we conceive a versatile saddle point approximation based method. Specifically, we first characterize the distribution of one user's uplink delay by the saddle point approximation method. Based on the characteristics of the synchronous and asynchronous downlink transmission schemes, the distributions of one iteration delay are further obtained along with the distributions of the overall delay. Besides, extreme value theory (EVT) \cite{extremevalue} and large deviation theory (LDT) \cite{LDT} are also exploited to reveal the asymptotic properties of the delay distribution in wireless FL systems. To address the challenge in theoretically deriving the accurate number of convergence rounds, we model the number of convergence rounds as a random variable in lieu of a constant, followed by using an empirical distribution to investigate the overall delay in the general wireless FL systems with nonconvex FL models.

The main contribution of this paper is the analysis of the distribution of the one iteration and overall delay in wireless FL systems. To the best of our knowledge, this is the first work that provides a theoretical analysis of the delay distribution in wireless FL systems. The major contributions are summarized as follows:
\begin{itemize}
\item We establish a unified FL modeling framework over wireless fading channels to analyze the distribution of training delay caused by wireless communication under the practical condition that the transmission time is larger than the coherence time.
\item A transmission model consisting of the uplink and downlink transmission is proposed, for which both synchronous and asynchronous downlink transmission schemes are developed for different application scenarios. 
\item To characterize the delay distribution in wireless FL systems, we propose a saddle point approximation based method. Moreover, EVT and LDT are leveraged to reveal asymptotic properties of the delay distribution in wireless FL systems.

\item Extensive experiments demonstrate that the simulation results are in good agreement with the established theoretical results. In particular, for one iteration delay, the theoretical results perfectly match with the simulation results, which verifies the validity of the proposed methods. 
\end{itemize}

The rest of this paper is organized as follow. Section II presents the system model, which contains the FL model, transmission and computation model, with the definition of delay in two different transmission schemes. Based on the FL model and transmission model, one iteration and overall delay analysis are provided in Section III. In Section IV, experiment settings and simulation results are described. The conclusion is given in Section V.

\section{System Model}
\vspace{-1mm}
As shown in Fig. 1, we consider a FL system over wireless channels that consists of one base station (BS) and a set $\mathcal{K}$ of $K$ users with all of them equipped with a single antenna. In this system, users and the BS perform the FL algorithm collaboratively to complete the data computation and analysis tasks. Specifically, a shared global model is trained by a decentralized machine learning method without transfering users' private data. In this section, we first explain the FL principle by introducing the canonical FL algorithm. The transmission and computation model, supported by synchronous and asynchronous communication schemes, are further established for the investigation of the delay distribution analysis.
\begin{figure}[t]
\centerline{\includegraphics[width=15cm]{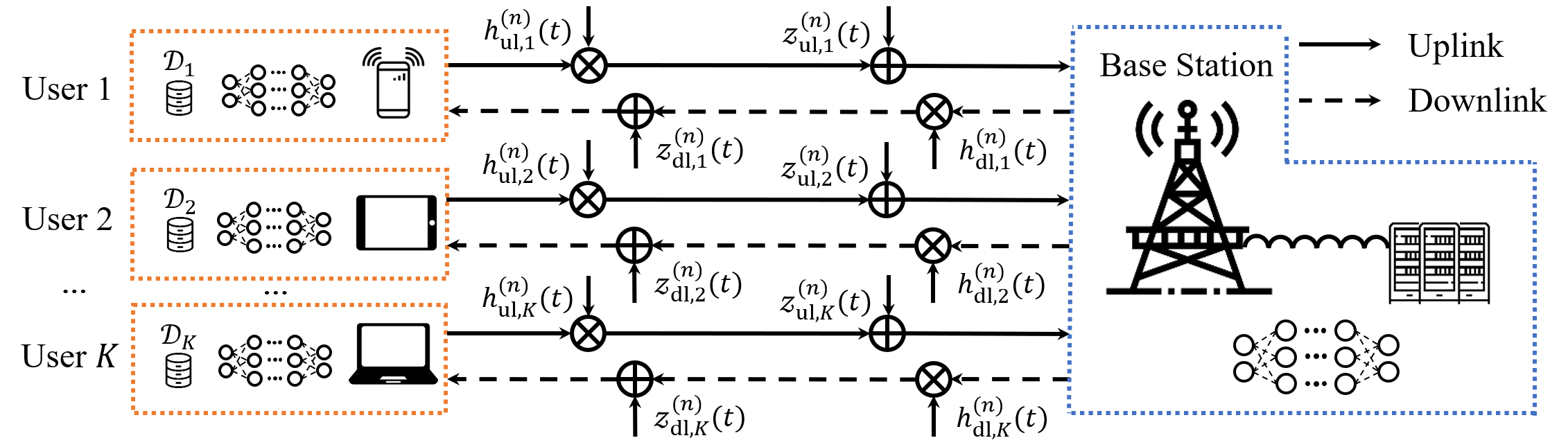}}
\vspace{-3mm}
\caption{Wireless federated learning system.}
\label{fig}
\vspace{-3mm}
\end{figure}
\vspace{-6mm}
\subsection{Federated Learning Model}
\vspace{-1mm}
Consider a wireless FL system as shown in Fig. 1. Let $\mathcal{D}_k=\{\boldsymbol{x}_{k,i},y_{k,i}\}_{i=1}^{|\mathcal{D}_k|}$ denote the local dataset of user $k$ with $|\mathcal{D}_k|$ samples, where $\boldsymbol{x}_{k,i}$ is the $i$-th training data sample with $y_{k,i}$ as the corresponding output.  The parameters of the shared  global model are denoted by $\boldsymbol w \in \mathbb{R}^m$. We then define $l(\boldsymbol w,\boldsymbol x_{k,i},y_{k,i})$ as the loss function to measure the learning performance. Various learning tasks and structures may yield different loss functions.
By this means, the local loss function of user $k$ can be defined as \cite{4}
\addtolength{\abovedisplayskip}{-2mm}
\addtolength{\belowdisplayskip}{-2mm}
\vspace{-0.9mm}
\begin{equation}
L_k(\boldsymbol w)=\frac{1}{|\mathcal{D}_{k}|}\sum_{i=1}^{|\mathcal{D}_{k}|}l(\boldsymbol w,\boldsymbol x_{k,i},y_{k,i}).\label{eq1}
\vspace{-0.5mm}
\end{equation}
The goal of the FL training process is to find a global model that minimizes the weighted sum of every involved user's loss functions. Therefore, the training procedure can be done by solving the following optimization problem:
\vspace{-1.3mm}
\begin{equation}
\underset{\boldsymbol w}{\min} \, L(\boldsymbol{w})=\sum_{k=1}^{K}\frac{|\mathcal{D}_k|}{D}L_k(\boldsymbol{w})=\frac{1}{D}\sum_{k=1}^K \sum_{i=1}^{|\mathcal{D}_k|}l(\boldsymbol{w},\boldsymbol{x}_{k,i},y_{k,i}),\label{eq2}
\vspace{-1.3mm}
\end{equation}
where $D=\sum_{k=1}^{K} |\mathcal{D}_{k}|$ is the total number of data samples from all users. Different FL algorithms can be applied to solve \eqref{eq2}. Such solution basically consists of two main steps, i.e., local update and global aggregation. Specifically, the local update is the process in which learning tasks are computed based on local datasets, while the global aggregation is achieved by updating the global model using the uploaded users' local model updates, followed by broadcasting the global parameters to them. This procedure repeats until convergence. We shall adopt the FL algorithm in \cite{algorithm}, \cite{4} and \cite{DANE} as presented in Alg. 1. Specifically, the global FL parameter at iteration $n$ is denoted by $\boldsymbol{w}^{(n)}$. After the broadcast of gradient $\nabla L(\boldsymbol{w}^{(n)})$, the local update can be obtained by solving  the following local FL problem:
\addtolength{\textfloatsep}{-9mm}
\begin{algorithm}[t]
\caption{Federated Learning Algorithm} 
\label{alg1}
{\small{
\begin{algorithmic}[1]
\STATE Initialize global parameters vector $\boldsymbol{w}^{(0)}$ and global iteration number $n=0$.
\REPEAT
\STATE Each user $k$ computes  $\nabla L_k(\boldsymbol{w}^{(n)})$ and sends it to the BS.
\STATE The BS computes 
\vspace{-2mm}
\begin{equation}
    \nabla L(\boldsymbol{w}^{(n)})=\frac{1}{K}\sum_{k=1}^K \nabla L_k(\boldsymbol{w}^{(n)}),\label{eq3}
\vspace{-2mm}
\end{equation}
and broadcasts it to all involved users.
\FOR{each user $k \in \mathcal{K}$ in parallel}
\STATE Solve local problem (5) to get the solution $\boldsymbol{g}_k^{(n)}$. 
\STATE Send $\boldsymbol{g}_k^{(n)}$ to the BS.
\ENDFOR
\STATE The BS computes 
\vspace{-2mm}
\begin{equation}
    \boldsymbol{w}^{(n+1)}= \boldsymbol{w}^{(n)}+\frac{1}{K}\sum_{k=1}^K \boldsymbol{g}_k^{(n)},\label{eq4}
\vspace{-3mm}
\end{equation}
and broadcasts it to all involved users.
\STATE Set $n=n+1$.
\UNTIL{the termination condition is satisfied.}
\end{algorithmic}}}
\end{algorithm}
\vspace{-1.5mm}
\begin{equation}
\underset{\boldsymbol{g}_k}{\min}  \  G_k (\boldsymbol{w}^{(n)},\boldsymbol{g}_{k}) \triangleq  L_k(\boldsymbol{w}^{(n)}+\boldsymbol{g}_k)-\big(\nabla L_k(\boldsymbol{w}^{(n)})-\xi \nabla L(\boldsymbol{w}^{(n)})\big)^T \boldsymbol{g}_k+\frac{\mu}{2}\|\boldsymbol{g}_k\|^2, \label{eq5}
\vspace{-0.5mm}
\end{equation}
where $\xi$ is a step size parameter, $\mu$ is a regularized parameter, $\boldsymbol{g}_k$ is the difference between the global FL model parameters and the local FL model parameters of user $k$. The local model parameters of user $k$ at iteration $n$ can be updated as $\boldsymbol{w}_{k}^{(n)}\!=\boldsymbol{w}^{(n)}\!+\boldsymbol{g}_{k}$. 
To establish convergence, the solution $\boldsymbol{g}_{k}^{(n)}$ at global iteration $n$ with the target accuracy $\eta$ needs to satisfy the requirement 
\vspace{-7.5mm}
\begin{equation}
G_k(\boldsymbol{w}^{(n)},\boldsymbol{g}_k^{(n)})-G_k(\boldsymbol{w}^{(n)},\boldsymbol{g}_{k}^{(n)\star}) \leq \eta \big(G_k(\boldsymbol{w}^{(n)},\boldsymbol{0})-G_k(\boldsymbol{w}^{(n)},\boldsymbol{g}_{k}^{(n)\star})\big),\label{eq6}
\vspace{-4mm}
\end{equation}
where $\boldsymbol{g}_{k}^{(n)\star}$ is the optimal solution of problem (5). After computing $\boldsymbol{g}_{k}^{(n)}$ based on the local dataset, each user sends this result, instead of raw data, to the BS to carry on \eqref{eq4} without leaking private information. Then the BS broadcasts the global model update $\boldsymbol{w}^{(n+1)}$ to the involved users for a new learning iteration. Similar to the local update, for Problem \eqref{eq2}, to achieve a feasible solution $\boldsymbol w^{(n)}$ under a given accuracy $\epsilon_{0}$ by iterating the local update and global aggregation process, the termination condition for Alg. 1 can be expressed as
\vspace{-2mm}
\begin{equation}
L(\boldsymbol{w}^{(n)})-L(\boldsymbol{w}^\star) \leq \epsilon_{0}\big(L(\boldsymbol{w}^{(0)})-L(\boldsymbol{w}^\star)\big),\label{eq7}
\vspace{-2mm}
\end{equation}
where $\boldsymbol{w}^\star$ is the optimal solution of Problem \eqref{eq2}.
\vspace{-6mm}
\subsection{Transmission and Computation Model}
\vspace{-0.5mm}
From above discussions on FL, we can see that communication between end users and BS is critical for local and global model updates. Specifically, users upload their local model parameters to the BS via uplink transmission for global model update, while BS broadcasts the updated global model parameters to all users via downlink transmission. Due to the limited resources and deep wireless channel fading, the transmission procedure can not be accomplished without delay. In this subsection, we shall present the transmission and computation model in details. 

Consider a block fading channel model, in which the channel coefficient remains constant within a time slot of length $T_0$ and varies in an independent and identically distributed (i.i.d.) manner across slots and users. We consider a narrowband scenario where the transmission time of the learning procedure is much longer than the channel coherence time $T_0$. This scenario is practical because the bandwidth and the transmission power are very limited especially for the Internet of Things (IoT) or Industrial Internet applications. 

For the uplink transmission, we assume that each BS-user pair transmits independently via frequency domain multiple access (FDMA) \cite{fdma}. Then for input ${x}_{{\rm ul},k}^{(n)}(t)\in \mathbb{C}$ (i.e., the representation of local model update) of user $k$ at the $t$-th time slot of the uplink transmission in iteration $n$, the corresponding output $y_{{\rm ul},k}^{(n)}(t) \in \mathbb{C}$ is given by
\vspace{-1mm}
\begin{equation}
y_{{\rm ul},k}^{(n)}(t)={h}_{{\rm ul},k}^{(n)}(t) {x}_{{\rm ul},k}^{(n)}(t) + z_{{\rm ul},k}^{(n)}(t),\label{eq8}
\vspace{-1.5mm}
\end{equation}
where ${h}_{{\rm ul},k}^{(n)}(t) \in \mathbb{C}$ is the uplink Rayleigh fading channel coefficient from user $k$ to the BS, $z_{{\rm ul},k}^{(n)}(t) \in \mathbb{C}$ is the additive Gaussian noise.

Moreover, we assume that each dimension of the uploading parameters is quantified by $q$ nats$\footnote{For the convenience of derivation, we use nats instead of bits to denote the size of data. }$. Thus, one user needs to upload $S=mq$ nats at iteration $n$. Let $r_{{\rm ul},k}^{(n)}(t)$ (in nats/s) denote the uplink achievable data rate of user $k$ during the $t$-th uplink transmission time slot in iteration $n$, then the uplink transmission delay of user $k$ at iteration $n$ is given by
\vspace{-1.5mm}
\begin{equation}
t_{{\rm ul},k}^{(n)}= T_0\min\,  \left \{d:T_0\sum_{t=1}^{d}r_{{\rm ul},k}^{(n)}(t) \geq S, d \in \mathbb{N}_{+} \right \}.\label{eq9}
\vspace{-1mm}
\end{equation}

Similarly, for the downlink transmission, input is the aggregation results $x_{\rm dl}^{(n)}(t) \in \mathbb{C}$ from the BS, and the corresponding output $y_{{\rm dl},k}^{(n)}(t) \in \mathbb{C}$ of user $k$ at the $t$-th time slot of downlink transmission in iteration $n$ is given by
\vspace{-1mm}
\begin{equation}
y_{{\rm dl},k}^{(n)}(t)= h_{{\rm dl},k}^{(n)}(t) x_{{\rm dl}}^{(n)}(t)+ z_{{\rm dl},k}^{(n)}(t),\label{eq10}
\vspace{-2mm}
\end{equation} 
where $h_{{\rm dl},k}^{(n)}(t) \in \mathbb{C}$ is the downlink Rayleigh fading channel coefficient from the BS to user $k$, $z_{{\rm dl},k}^{(n)}(t) \in \mathbb{C}$ is the additive Gaussian noise. Let $r_{{\rm dl},k}^{(n)}(t)$ (in nats/s) denote the downlink achievable data rate of user $k$ during the $t$-th downlink transmission time slot in iteration $n$, then the downlink transmission delay of user $k$ at iteration $n$ is given by
\begin{equation}
t_{{\rm dl},k}^{(n)}= T_0\min\, \left \{d:T_0\sum_{t=1}^{d}r_{{\rm dl},k}^{(n)}(t) \geq S, d \in \mathbb{N}_{+}\right \}.\label{eq11}
\end{equation}

For the computation time consumption, due to the relatively stronger computational capability of the BS, we ignore the model aggregation delay at the BS. Thus, the computation time consumption at iteration $n$ is mainly from the local computation latency, which is defined as $t_{{\rm cp},k}^{(n)}=C_k |\mathcal{D}_k|$ with $C_k$ being a constant representing the computational capability of device $k$. This definition is widely adopted in the literature for delay minimization in FL \cite{huangkb2,algorithm,Chenmz}. 

The global model aggregation will not start until all involved users' model parameters arrived at the BS. Therefore, based on $t_{{\rm ul},k}^{(n)}$, $t_{{\rm dl},k}^{(n)}$ and $t_{{\rm cp},k}^{(n)}$, the one learning iteration delay $T^{(n)}$ at iteration $n$ is given by$\footnote{Actually, in Alg. \ref{alg1}, each iteration consists of two rounds of computation and transmission: one round for $\nabla L_k(\boldsymbol{w}^{(n)})$ and $\nabla L(\boldsymbol{w}^{(n)})$, another round for $\boldsymbol{g}_k^{(n)}$ and $\boldsymbol{w}^{(n+1)} $. Without loss of generality, we consider only one round of computation and transmission for the convenience of analysis. Also note that, in this paper, the one iteration delay is denoted by $T^{(n)}$ when it is used as a variable. It can also be written as a function of the number of involved users $K$ in the FL system, i.e., $T^{(n)}(K)$.}$
\vspace{-2mm}
\begin{equation}
T^{(n)}=\,\underset{k \in \mathcal{K}}{\max}\, \{t_{{\rm ul},k}^{(n)}+t_{{\rm dl},k}^{(n)}+t_{{\rm cp},k}^{(n)}\}.\label{eq12}
\vspace{-2.5mm}
\end{equation} 

Given the number of iterations for convergence $N$, the overall delay $T_c$ is given by
\vspace{-1mm}
\begin{equation}
T_c=\sum_{i=0}^{N-1}T^{(i)}.
\vspace{-2mm}
\end{equation}

\subsection{Delay in Different Downlink Schemes}
In this subsection, we will specify the one iteration and overall delay of two different transmission schemes in wireless FL systems. 
According to Shannon formula, the achievable uplink data rate $r_{{\rm ul},k}^{(n)}(t)$ of user $k$ at the $t$-th uplink transmission time slot in iteration $n$ is given by
\begin{equation}
r_{{\rm ul},k}^{(n)}(t)=B \ln \left(1+\frac{P_k|h_{{\rm ul},k}^{(n)}(t)|^2}{\sigma ^2}\right),\label{eq13}
\end{equation}
where $P_k$ is the transmission power of user $k$, $\sigma^2$ is the noise power, $B$ is the bandwidth. For the downlink transmission, we propose two different designs, i.e., the synchronous downlink scheme as the first one, while the asynchronous downlink scheme as the second one.

\subsubsection{Synchronous Downlink Scheme}
\quad

The first proposed transmission scheme is the synchronous downlink communication, for 
\addtolength{\textfloatsep}{5mm}
\begin{figure}[t]
\centerline{\includegraphics[width=16cm]{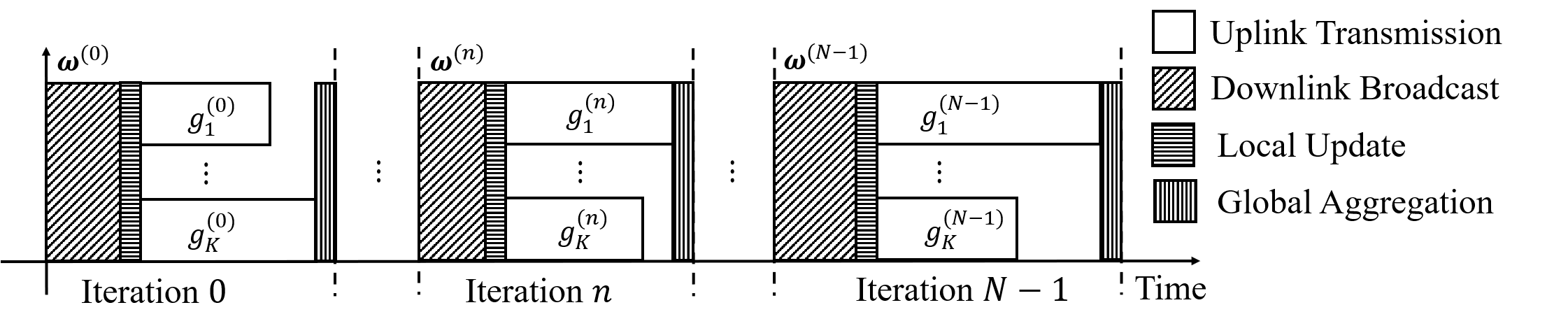}}
\vspace{-1mm}
\caption{The schematic diagram for the synchronous downlink scheme.}
\label{fig}
\vspace{-3mm}
\end{figure} 
which the downlink delay depends on the worst channel of involved users. Since the BS occupies more bandwidth to broadcast the global model, the bandwidth for downlink transmission is denoted by $B_{\rm dl}$. The downlink transmission rate at the $t$-th downlink transmission time slot is thus given by
\vspace{-0.5mm}
\begin{equation}
r_{{\rm dl}}^{(n)}(t)=B_{\rm dl} \ln \left(1+\frac{P_{\rm dl}\big(h_{{\rm dl}}^{(n)}(t)\big)^2}{\sigma ^2}\right),\label{eq14}
\vspace{-1mm}
\end{equation}
where $h_{{\rm dl}}^{(n)}(t)=\min \{|h_{{\rm dl},k}^{(n)}(t)|, k \in \mathcal{K}\}$, $P_{\rm dl}$ is the transmission power of the BS. In this scheme, users have the same downlink transmission delay. Therefore, all involved users update their local models synchronously. The schematic diagram for this scheme is shown in Fig. 2. Accordingly, the uplink and downlink delays at iteration $n$ in the first scheme are respectively given by
\begin{equation}
T_{\rm ul}^{(n)}=\max\{t_{{\rm ul},k}^{(n)},k \in \mathcal{K}\}=T_0\,\underset{k \in \mathcal{K}}{\max} \,  \min \left \{ d: T_0 \sum_{t=1}^{d}r_{{\rm ul},k}^{(n)}(t) \geq S, d \in \mathbb{N}_+\right \},\label{eq15}
\end{equation}

\begin{equation}
\quad\quad\quad\quad\quad\quad\,\, T_{\rm dl}^{(n)}= T_0\min \left \{ d: T_0 \sum_{t=1}^{d}r_{{\rm dl}}^{(n)}(t) \geq S, d \in \mathbb{N}_+ \right \}.\label{eq16}
\vspace{-3mm}
\end{equation}  

As the local computation time $t_{{\rm cp},k}^{(n)}$ is a deterministic constant for user $k$, it has no influence on the random distribution of the one iteration delay. Moreover, with the rapid development of both algorithms and hardware, the computational power and efficiency of mobile devices is growing rapidly. Thus, for simplicity, it is reasonable to assume that $t_{{\rm cp},k}^{(n)}$ is much shorter than the uplink and downlink transmission delay and can be ignored in the following discussions$\footnote{The reasonability of this assumption will be further proved in Section \textrm{IV}.}$.

The one iteration delay at iteration $n$ in this scheme is thus given by $T^{(n)}=T_{\rm ul}^{(n)}+T_{\rm dl}^{(n)}$. From the above analysis, we can see that in the synchronous downlink scheme there are two time alignments among all users during one iteration. The first one is downlink time alignment because of the synchronous downlink scheme, while the second one is one iteration time alignment because the global model aggregation can not start until all users' local model parameters are uploaded to the BS.
\subsubsection{Asynchronous Downlink Scheme}
\quad 
\begin{figure}[t]
\centerline{\includegraphics[width=15cm]{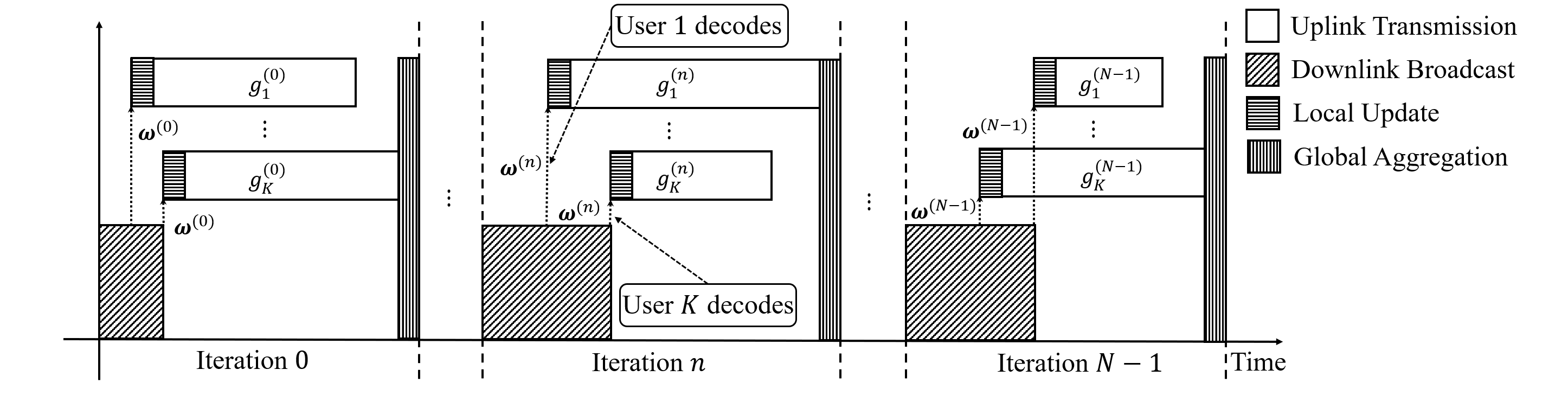}}
\vspace{-3mm}
\caption{The schematic diagram for the asynchronous downlink scheme.}
\label{fig}
\vspace{-3mm}
\end{figure}

Another proposed transmission scheme is the asynchronous downlink communication, which is also known as broadcast with common information. This downlink model is equivalent to a broadcast channel in which a single encoding of a common message is being sent to multiple receivers, where each experiences a different SNR because of different channel coefficients. By using rateless coding, users can achieve their capacity simultaneously \cite{rateless}.  Therefore, in this case, the downlink transmission rate for user $k$ is given by 
\begin{equation}
r_{{\rm dl},k}^{(n)}(t)=B_{\rm dl} \ln \left (1+\frac{P_{\rm dl}|h_{{\rm dl},k}^{(n)}(t)|^2}{\sigma ^2}\right ).\label{eq17}
\end{equation}

In this scheme, the start time of different users' uplink transmission is asynchronous. Therefore, as shown in Fig. 3, there is only one time alignment between all users in one iteration. Hence, one iteration delay in this scheme is given by
\vspace{-1mm}
\begin{align}
T^{(n)}&=\underset{k \in \mathcal{K}}{\max} \,\{T_{k}^{(n)}:T_{k}^{(n)}=t_{{\rm ul},k}^{(n)}+t_{{\rm dl},k}^{(n)}\}\nonumber \\
&=T_0\, \underset{k \in \mathcal{K}}{\max}\, \min \left \{d_1+d_2: T_0\sum_{t=1}^{d_1}r_{{\rm dl},k}^{(n)}(t) \geq S, T_0\sum_{t=1}^{d_2} r_{{\rm ul},k}^{(n)}(t) \geq S; d_1, d_2 \in \mathbb{N}_{+} \right \},\label{eq18}
\end{align}
which is the largest sum of the uplink and downlink transmission delay.
\vspace{-2mm}
\section{Delay analysis in wireless FL systems}
In this section, we first analyze the distribution of one iteration delay in two proposed wireless FL transmission schemes by leveraging a saddle point approximation based method \cite{SPA}. Based on an upper bound of the iteration numbers for convergence, we also analyze the overall delay distribution. To get more insights, we use EVT, LDT and stochastic order \cite{ross} to further characterize the properties of the delay distribution in wireless FL systems. Finally, to overcome the difficulties of obtaining the accurate number of convergence rounds analytically, we model the number of convergence rounds as a random variable, thereby using an empirical distribution to further characterize the overall delay in wireless FL systems.
\subsection{One Iteration Delay Distribution of the Synchronous Downlink Scheme}
For the synchronous downlink scheme, we investigate the uplink delay and downlink delay presented in Section \textrm{II}. To derive the uplink delay distribution of the wireless FL system, we conceive a saddle point approximation based method, in which the Lugannani-Rice (LR) formula \cite{LR} and differential analysis are leveraged to get one user's uplink delay distribution in Lemma 1, followed by presenting the distribution of the system's uplink delay in Theorem 1. For simplicity, we assume that the mean of the channel coefficients is $\sqrt{\pi / 2}$. Moreover, we assume that the transmit power of all involved users is fixed, for which we let $\lambda=\frac{P_k}{\sigma^2}$ and $\lambda_d=\frac{P_{\rm dl}}{\sigma^2}$ in the following derivations.

\textbf{Lemma 1.}
Define $Z_d$ as
\begin{equation}
Z_d=
\begin{cases}
\frac{S}{BT_0} , & \quad \text{$d = 0$,}\\
\frac{S}{BT_0}-\sum_{t=1}^{d}\frac{r_{{\rm ul},k}^{(n)}(t)}{B},& \quad \text{$d \geq 1$.}
\end{cases}\label{zd}
\vspace{1mm}
\end{equation}
For all $d \geq 1$, the distribution of user $k$'s uplink delay $t_{{\rm ul},k}^{(n)}$ can be expressed as
\vspace{1mm}
\begin{equation}
\varrho(d)\!=\!\Pr\big\{t_{{\rm ul},k}^{(n)}\!=\! dT_0\big\}\!=\! \frac{1}{\sqrt{2\pi}} \bigg(\! \int_{\omega_{d-1}}^{\omega_{d}} e^{-\frac{u^2}{2}}du+e^{-\frac{\omega_{d-1}^2}{2}}\Big(\frac{1}{\psi_{d-1}}-\frac{1}{\omega_{d-1}}\Big)-e^{-\frac{\omega_{d}^2}{2}}\Big(\frac{1}{\psi_d}-\frac{1}{\omega_d}\Big) \! \bigg),\label{eq20}
\vspace{2mm}
\end{equation}
where $\omega_d={\rm sign}\big(s^*_d\big)\sqrt{-2K_d \big(s^*_d\big)}$, $\psi_d=s^*_d\sqrt{K_d''\big(s^*_d\big)} $ for $d \geq 1$, and $\omega_0=\psi_0=-\infty$. $K_{d}(s)$ is the Cumulant Generating Function (CGF) of $Z_d$, which is given in \eqref{lemma1a1}. Here, $s^*_d$ is the solution to $K_d' \big(s^*_d\big)=0$, which satisfies \eqref{lemma1a2}.
\begin{align}
    K_d(s)=\frac{S}{BT_0}s+d\ln\left(e^{\frac{1}{2\lambda}}(2\lambda)^{-s}\Gamma\Big(-s+1,\frac{e^{r_{\rm th}}}{2\lambda}\Big)\right). \label{lemma1a1}
\end{align}
\begin{equation}
\frac{S}{BT_0}=\frac{dG_{2,3}^{3,0}\Bigg(\begin{array}{c}0,0\\-s^*_d,-1,-1\end{array}\Bigg\vert \frac{1}{2\lambda}\Bigg)}{2\lambda \Gamma(-s^*_d+1,\frac{1}{2\lambda})}.\label{lemma1a2}
\vspace{-1mm}
\end{equation}
In \eqref{lemma1a1}-\eqref{lemma1a2}, $\Gamma(s,x)\triangleq \int_x^{+\infty}t^{s-1}e^{-t}dt$ is the upper incomplete gamma function \cite{gammafunc} and
\begin{equation}
G_{\rho_3,\rho_4}^{\rho_1,\rho_2}\Bigg(\begin{matrix}a_1,a_2,\ldots,a_{\rho_3}\\b_1,b_2,\ldots,b_{\rho_4} \end{matrix}\Bigg|z\Bigg) =\frac{1}{2\pi i}\oint_{\mathcal{L}}\frac{\prod_{j=1}^{\rho_1}\Gamma(b_j-h)\prod_{j=2}^{\rho_2}\Gamma(1-a_j+h)}{\prod_{j=\rho_1+1}^{\rho_4}\Gamma(1-b_j+h)\prod_{j=\rho_2+1}^{\rho_3}\Gamma(a_j-h)}z^h dh\label{gfunc}
\vspace{1mm}
\end{equation}
is the Meijer G-function \cite{MGfunc}.
\begin{IEEEproof}
Please refer to Appendix A for details.
\end{IEEEproof}

\textbf{Theorem 1.}
Given the distribution of $t_{{\rm ul},k}^{(n)}$, the system's uplink delay distribution in the synchronous downlink scheme is given by
\vspace{-5mm}
\begin{align}
 \varphi(d)=   \Pr\big\{ T_{\rm ul}^{(n)}=dT_0\big\}&=\bigg(\frac{1}{\sqrt{2\pi}}\int_{-\infty}^{\omega_d}e^{-\frac{u^2}{2}}du-\frac{e^{-\frac{\omega_d^2}{2}}}{\sqrt{2\pi}}\Big(\frac{1}{\psi_d}-\frac{1}{\omega_d}\Big)\bigg)^K \nonumber \\
&-\bigg(\frac{1}{\sqrt{2\pi}}\int_{-\infty}^{\omega_{d-1}}e^{-\frac{u^2}{2}}du-\frac{e^{-\frac{\omega_{d-1}^2}{2}}}{\sqrt{2\pi}}\Big(\frac{1}{\psi_{d-1}}-\frac{1}{\omega_{d-1}}\Big)\bigg)^K.\label{eq21} 
\end{align}

\vspace{3mm}
\begin{IEEEproof}
Please refer to Appendix B for details.
\end{IEEEproof}

The distribution of the downlink delay in the synchronous downlink scheme can be derived similarly from the uplink delay analysis. It is given in the following lemma.

\textbf{Lemma 2.} Define $\tilde Z_d$ as
\begin{equation}
\tilde Z_d=
\begin{cases}
\frac{S}{B_{\rm dl}T_0} , & \quad \text{$d = 0$,}\\
\frac{S}{B_{\rm dl}T_0}-\sum_{t=1}^{d}\frac{r_{{\rm dl}}^{(n)}(t)}{B_{\rm dl}},& \quad \text{$d \geq 1$.}
\end{cases}\label{zddl}
\vspace{1mm}
\end{equation}
For all $d \geq 1$, the downlink delay distribution in the synchronous downlink scheme is given by
\vspace{1mm}
\begin{equation}
\upsilon(d)\!=\! \Pr\big\{T_{\rm dl}^{(n)}\!=\!dT_0\big\}\!=\!\frac{1}{\sqrt{2\pi}}\bigg(\! \int_{\tilde \omega_{d-1}}^{\tilde \omega_{d}}\! e^{-\frac{u^2}{2}}du+e^{-\frac{\tilde \omega_{d-1}^2}{2}}\Big(\frac{1}{\tilde \psi_{d-1}}\!-\! \frac{1}{\tilde \omega_{d-1}}\Big)\!-\!e^{-\frac{\tilde \omega_{d}^2}{2}}\Big(\frac{1}{\tilde \psi_d}\!-\!\frac{1}{\tilde \omega_d}\Big) \! \bigg),\label{lemma3.1}
\vspace{2mm}
\end{equation}
where $\tilde\omega_d \!=\!{\rm sign}\big(\tilde s^*_d\big)\sqrt{-2\tilde K_d \big(\tilde s^*_d\big)}$, $\tilde \psi_d\!=\!\tilde s^*_d\sqrt{\tilde K_d''\big(\tilde s^*_d\big)} $ for $d\geq1$, and $\tilde\omega_0\!=\!\tilde\psi_0\!=\!-\infty$. $\tilde K_d( s)$ is the CGF of $\tilde Z_d$, which is given in \eqref{lemma3.2}. Here, $\tilde s^*_d$ is the solution to $\tilde K_d' \big(\tilde s^*_d\big)=0$, which satisfies \eqref{lemma3.3}.
\vspace{-4mm}
\begin{equation}
    \tilde K_d(s)=\frac{S}{B_{\rm dl}T_0}s+d\ln\left(e^{\frac{K}{2\lambda_d}}\Big(\frac{2\lambda_d}{K}\Big)^{-s}\Gamma\Big(-s+1,\frac{K}{2\lambda_d}\Big)\right). \label{lemma3.2}
\end{equation}
\begin{equation}
\frac{S}{B_{\rm dl}T_0}=\frac{dKG_{2,3}^{3,0}\Bigg(\begin{array}{c}0,0\\-\tilde s^*_d,-1,-1\end{array}\Bigg\vert \frac{K}{2\lambda_d}\Bigg)}{2\lambda_d \Gamma(-\tilde s^*_d+1,\frac{K}{2\lambda_d})}.
\label{lemma3.3}
\end{equation}
\begin{IEEEproof}
Please refer to Appendix C for details.
\end{IEEEproof}
Based on Theorem 1 and Lemma 2, we are ready to obtain the distribution of one iteration delay in the synchronous downlink scheme in the following theorem.

\textbf{Theorem 2.}
Given the uplink and downlink delay distribution, the distribution of one iteration delay in the synchronous downlink scheme can be obtained as
\begin{align}
    &\Pr\big\{ T^{(n)}=dT_0\big\} =\sum_{i=1}^{d-1} \varphi(i) \, \upsilon(d-i) 
\nonumber\\ 
&=\frac{1}{\sqrt{2\pi}}\sum_{i=1}^{d-1}\bigg[\bigg(\frac{1}{\sqrt{2\pi}}\int_{-\infty}^{\omega_i}e^{-\frac{u^2}{2}}du-\frac{e^{-\frac{\omega_i^2}{2}}}{\sqrt{2\pi}}\Big(\frac{1}{\psi_i}-\frac{1}{\omega_i}\Big)\bigg)^K \!- \! \bigg(\frac{1}{\sqrt{2\pi}}\int_{-\infty}^{\omega_{i-1}}e^{-\frac{u^2}{2}}du -\frac{e^{-\frac{\omega_{i-1}^2}{2}}}{\sqrt{2\pi}} \nonumber\\
&\Big(\!\frac{1}{\psi_{i-1}}\! -\! \frac{1}{\omega_{i-1}}\!\Big)\!\bigg)^{\!\!K} \bigg]    \bigg[\! \int_{\tilde \omega_{d-i-1}}^{\tilde \omega_{d-i}} \!\!\! e^{-\frac{u^2}{2}}du+ \! e^{-\frac{\tilde \omega_{d-i-1}^2}{2}}\Big(\frac{1}{\tilde \psi_{d-i-1}}\! - \! \frac{1}{\tilde \omega_{d-i-1}}\Big)\!-\!e^{-\frac{\tilde \omega_{d-i}^2}{2}}\Big(\frac{1}{\tilde \psi_{d-i}}\!- \! \frac{1}{\tilde \omega_{d-i}}\Big) \bigg].\label{eq25}
\end{align}
\begin{IEEEproof}
As presented in Section \textrm{II-A}, in the synchronous downlink scheme, $T^{(n)}=T_{\rm ul}^{(n)}+T_{\rm dl}^{(n)}$. Therefore, given the uplink delay distribution and the downlink delay distribution, the distribution of one iteration delay $T^{(n)}$ is the convolution of distributions of $T_{\rm ul}^{(n)}$ and $T_{\rm dl}^{(n)}$, which is shown in \eqref{eq25}.
\end{IEEEproof}

\textbf{Corollary 1.}
When the number of users $K$ approaches infinity, define $R(t)$ as
\begin{equation}
R(t)=T_0\Big\lceil \frac{t}{T_0} \Big\rceil-t+\frac{\sum_{d=\lceil \frac{t}{T_0} \rceil}^{\infty}\big(1-\sum_{j=1}^{d}\sum_{i=1}^{j-1}\varrho(i)\upsilon(j-i)\big)}{1-\sum_{j=1}^{ \lfloor \frac{t}{T_0} \rfloor}\sum_{i=1}^{j-1}\varrho(i)\upsilon(j-i)} ,
\vspace{0.5mm}
\end{equation}
where $\lceil \cdot \rceil$ is the ceiling functions, $\lfloor \cdot \rfloor$ is the floor function. For all real $x$, if 
\begin{equation}
\lim_{t \to +\infty}\frac{1-\sum_{j=1}^{\lfloor \frac{t+xR(t)}{T_0} \rfloor}\sum_{i=1}^{j-1}\varrho(i)\upsilon(j-i)}{1-\sum_{j=1}^{\lfloor \frac{t}{T_0} \rfloor}\sum_{i=1}^{j-1}\varrho(i)\upsilon(j-i)}=e^{-x}, \label{coro1}
\vspace{0.5mm}
\end{equation}
then the limiting distribution of $T^{(n)}$ is given by
\vspace{-1.5mm}
\begin{equation}
\Pr \big \{T^{(n)}<y \big \}=e^{-e^{-\frac{y-a}{b}}},\label{co1}
\vspace{-4mm}
\end{equation}
where $a$ and $b$ can be chosen as
\vspace{-1.5mm}
\begin{equation}
a =\inf \bigg \{x: 1-\sum_{j=1}^{\lfloor \frac{x}{T_0} \rfloor}\sum_{i=1}^{j-1}\varrho(i)\upsilon(j-i)\leq \frac{1}{K}\bigg \},\label{coro1a}
\end{equation}
\begin{equation}
b =R(a).\label{coro1b}\quad \quad \quad \quad \quad \quad \quad \quad \quad \quad \quad \quad \quad \quad \,\,\,
\vspace{-4mm}
\end{equation}
\begin{IEEEproof}
Please refer to Appendix D for details.
\end{IEEEproof}
In Corollary 1, we use EVA to characterize the asymptotic property of one iteration delay in the synchronous downlink scheme. In practical scenarios, the number of users in a wireless system is usually large. Hence, we can use Corollary 1 to simplify the computation. Note that, this analysis is also applicable to the one iteration delay in the asynchronous downlink scheme. 

\vspace{-3mm}
\subsection{One Iteration Delay Distribution of the Asynchronous Downlink Scheme}
The analysis of one iteration delay for the asynchronous downlink scheme has some similarities with the synchronous one in Section \textrm{III-A}. First, we give the distribution of $t_{{\rm dl},k}^{(n)}$ in the asynchronous downlink scheme in Lemma 3. Furthermore, the distribution of one iteration delay in the asynchronous downlink scheme will be presented in Theorem 3.

\textbf{Lemma 3.} Define $\hat Z_d$ as
\vspace{-1mm}
\begin{equation}
\hat Z_d=
\begin{cases}
\frac{S}{B_{\rm dl}T_0} , & \quad \text{$d = 0$,}\\
\frac{S}{B_{\rm dl}T_0}-\sum_{t=1}^{d}\frac{r_{{\rm dl},k}^{(n)}(t)}{B_{\rm dl}},& \quad \text{$d \geq 1$.}
\end{cases}
\end{equation}
For all $d \geq 1$, the distribution of $t_{{\rm dl},k}^{(n)}$ in the asynchronous downlink scheme is given by
\begin{equation}
\vartheta(d)\!=\! \Pr\big\{t_{{\rm dl},k}^{(n)}\! =\! dT_0\big\}\!=\!\frac{1}{\sqrt{2\pi}}\bigg(\!\int_{\hat \omega_{d-1}}^{\hat \omega_{d}}\! e^{-\frac{u^2}{2}}du+e^{-\frac{\hat \omega_{d-1}^2}{2}}\Big(\frac{1}{\hat \psi_{d-1}}-\frac{1}{\hat \omega_{d-1}}\Big)-e^{-\frac{\hat \omega_{d}^2}{2}}\Big(\frac{1}{\hat \psi_d}-\frac{1}{\hat \omega_d}\Big) \! \bigg),\label{zhat}
\end{equation}
where $\hat \omega_d\!=\!{\rm sign}\big(\hat s^*_d\big)\sqrt{-2\hat K_d \big(\hat s^*_d\big)}$, $\hat \psi_d\!=\!\hat s^*_d\sqrt{\hat K_d''\big(\hat s^*_d\big)} $ for $d\geq1$, and $\hat \omega_0\!=\!\hat \psi_0\!=\!-\infty$. $\hat K_d(s)$ is the CGF of $\hat Z_d$, which is given in \eqref{lemma4.1}. Here, $\hat s^*_d$ is the solution to $\hat K_d' \big(\hat s^*_d\big)=0$, which satisfies \eqref{lemma4.2}.
\vspace{-6mm}
\begin{equation}
    \hat K_d(s)=\frac{S}{B_{\rm dl}T_0}s+d\ln\left(e^{\frac{1}{2\lambda_d}}(2\lambda_d)^{-s}\Gamma\Big(-s+1,\frac{1}{2\lambda_d}\Big)\right).\label{lemma4.1} 
\vspace{-3mm}
\end{equation}
\begin{equation}
\frac{S}{B_{\rm dl}T_0}=\frac{dG_{2,3}^{3,0}\Bigg(\begin{array}{c}0,0\\-\hat s^*_d,-1,-1\end{array}\Bigg\vert \frac{1}{2\lambda_d}\Bigg)}{2\lambda_d \Gamma(-\hat s^*_d+1,\frac{1}{2\lambda_d})}.\label{lemma4.2} 
\vspace{-2mm}
\end{equation}
\begin{IEEEproof}
This derivation is similar to Lemma 1. Except for the SNR varying from $\lambda$ to $\lambda_d$, the only difference is related to the bandwidth which becomes $B_{\rm dl}$. 
\end{IEEEproof}
\textbf{Theorem 3.}
The distribution of one iteration delay in the asynchronous downlink scheme is given by
\vspace{-2mm}
\begin{equation}
\Pr\big\{ T^{(n)}=dT_0 \big\}=\bigg(\sum_{j=1}^{d} \sum_{i=1}^{j-1}\varrho(i)\vartheta(j-i)\bigg)^K-\bigg(\sum_{j=1}^{d-1} \sum_{i=1}^{j-1}\varrho(i)\vartheta(j-i)\bigg)^K .\label{t2}
\end{equation}
\begin{IEEEproof}
Please refer to Appendix E for details.
\end{IEEEproof}

\textbf{Corollary 2.}
When the number of users $K$ approaches infinity, define $m(t)$ as
\begin{equation}
m(t)=T_0 \Big \lceil \frac{t}{T_0} \Big \rceil-t+\frac{\sum_{d=\lceil \frac{t}{T_0} \rceil}^{\infty}\big(1-\sum_{j=1}^{d}\sum_{i=1}^{j-1}\varrho(i)\vartheta(j-i)\big)}{1-\sum_{j=1}^{\lfloor \frac{t}{T_0} \rfloor}\sum_{i=1}^{j-1}\varrho(i)\vartheta(j-i)}.
\vspace{-1mm}
\end{equation}
For all real $x$, if 
\vspace{-1mm}
\begin{equation}
\lim_{t \to +\infty}\frac{1-\sum_{j=1}^{\lfloor \frac{t+xm(t)}{T_0} \rfloor}\sum_{i=1}^{j-1}\varrho(i)\vartheta(j-i)}{1-\sum_{j=1}^{\lfloor \frac{t}{T_0} \rfloor}\sum_{i=1}^{j-1}\varrho(i)\vartheta(j-i)}=e^{-x}
,\label{coro2}
\vspace{-3mm}
\end{equation}
then the limiting distribution of $T^{(n)}$ is given by
\vspace{-1.5mm}
\begin{equation}
\Pr \big\{T^{(n)}<y\big\}=e^{-e^{-\frac{y-a}{b}}}, \label{co2}
\vspace{-3.5mm}
\end{equation}
where $a$ and $b$ can be chosen as
\vspace{-1mm}
\begin{equation}
a=\inf \bigg \{x:1-\sum_{j=1}^{\lfloor \frac{x}{T_0} \rfloor}\sum_{i=1}^{j-1}\varrho(i)\vartheta(j-i)\leq \frac{1}{K} \bigg \}
,\label{coro2a}
\vspace{-1mm}
\end{equation}
\begin{equation}
b=m(a).\label{coro2b}\quad\quad\quad\quad\quad\,\,\quad\quad\quad\quad\quad\quad\quad\quad\quad
\vspace{-5mm}
\end{equation}
\begin{IEEEproof}
Please refer to Appendix F for details.
\end{IEEEproof}
\vspace{-5mm}
\subsection{Overall Delay Analysis}
It is challenging to drive the exact number of iterations for the distributed FL algorithms, for which we adopt the results in \cite{algorithm} presented in Lemma 4. It gives an upper bound of the iteration numbers needed for Alg. 1 to achieve convergence.

\textbf{Lemma 4.}
Assume that the loss function $L_k(\boldsymbol w)$ satisfies:
\begin{itemize}
\item[(a)] $L_k(\boldsymbol w)$ is $\alpha$-smooth, i.e., $\nabla^2 L_k(\boldsymbol w)\preceq \alpha \boldsymbol I$.
\item[(b)] $L_k(\boldsymbol w)$ is $\gamma$-strongly convex, i.e., $\nabla^2 L_k(\boldsymbol w) \succeq \gamma \boldsymbol I$. 
\end{itemize}

Under these assumptions, if we run Alg. 1 with $0<\xi \leq \frac{\gamma}{\alpha}$ and $\mu=0$ for 
\vspace{-1mm}
\begin{equation}
    n \geq \frac{u}{1-\eta}\triangleq I_0
\vspace{-0.5mm}
\end{equation}
iterations with $u=\frac{2\alpha^2}{\gamma ^2 \xi}\ln{\frac{1}{\epsilon _0}}$, we have $L(\boldsymbol{w}^{(n)})-L(\boldsymbol{w}^{\star})\leq \epsilon_0 \big(L(\boldsymbol{w}^{(0)})-L(\boldsymbol{w}^{\star})\big)$.
\begin{IEEEproof}
Please refer to \cite{algorithm} Appendix A for details.
\end{IEEEproof}

The assumptions in Theorem 3 are widely used in the literature for FL convergence analysis \cite{huangkb1,algorithm,JSAC2019,Chenmz,poor1,Niu}. 
Note that, Lemma 4 is established on the condition that $\mu=0$, which seems like abandoning the regularization term. However, if we put the regularization term into $L_k(\boldsymbol w)$, which means we define $L_k(\boldsymbol w)$ as the sum of the local loss and the regularization term, the conclusion in Lemma 4 still holds.

Having the upper bound of the number of iterations, we analyze the tail distribution of $T_c$ for different schemes by LDT in the following Theorems 4 and 5, respectively. 

\textbf{Theorem 4.}
For $\tau > I_0\mathbb{E}\{T^{(n)}\}$, if the CGF of $T^{(n)}$ exists, the tail distribution of $T_c$ in the synchronous downlink scheme can be expressed as
\vspace{-1mm}
\begin{equation}
\Pr\{ T_c \geq \tau\}=\exp\bigg(-s^\star \tau+I_0\ln \Big( \sum_{d=1}^{\infty} e^{s^\star dT_0} \sum_{i=1}^{d-1} \varphi(i)\upsilon(d-i)\Big)\bigg), \label{th4.1}
\vspace{-1mm}
\end{equation}
where $s^\star$ satisfies
\begin{equation}
\frac{\sum_{d=1}^{\infty} dT_0e^{s^\star dT_0} \sum_{i=1}^{d-1} \varphi(i)\upsilon(d-i)}{\sum_{d=1}^{\infty} e^{s^\star dT_0} \sum_{i=1}^{d-1} \varphi(i)\upsilon(d-i)}=\frac{\tau}{I_0}.\label{s*1}
\vspace{-1mm}
\end{equation}
\begin{IEEEproof}
Please refer to Appendix G for details.
\end{IEEEproof} 
By the same way, we can get the property of $T_c$'s tail distribution in the asynchronous downlink scheme in Theorem 5.

\textbf{Theorem 5.} For $\tau> I_0\mathbb{E}\{T^{(n)}\}$, if the CGF of $T^{(n)}$ exists, the tail distribution of $T_c$ in the asynchronous downlink scheme can be expressed as
\begin{equation}
\! \Pr\{ T_c \! \geq \! \tau\}\!=\!\exp \! \bigg(\! \! \!- \! s^\star \tau  +   I_0\ln\!\Big(\!\sum_{d=1}^{\infty} \!e^{s^\star dT_0} \big[\big(\!\sum_{j=1}^{d}\! \sum_{i=1}^{j-1}\varrho(i)\vartheta(j\! -\! i)\big)^K \! -  \big(\!\sum_{j=1}^{d-1} \! \sum_{i=1}^{j-1}\varrho(i)\vartheta(j\! -\! i)\big)^K  \big]  \!\Big)\! \!\bigg),
\vspace{-1mm}
\end{equation}
where $s^\star$ satisfies
\vspace{0.5mm}
\begin{equation}
\frac{\sum_{d=1}^{\infty} dT_0 e^{s^\star dT_0} \big[\big(\sum_{j=1}^{d} \sum_{i=1}^{j-1}\varrho(i)\vartheta(j-i))^K-(\sum_{j=1}^{d-1} \sum_{i=1}^{j-1}\varrho(i)\vartheta(j-i)\big)^K \big]}{\sum_{d=1}^{\infty} e^{s^\star dT_0} \big[(\sum_{j=1}^{d} \sum_{i=1}^{j-1}\varrho(i)\vartheta(j-i))^K-(\sum_{j=1}^{d-1} \sum_{i=1}^{j-1}\varrho(i)\vartheta(j-i)\big)^K \big]}=\frac{\tau}{I_0}.\label{s*2}
\vspace{1mm}
\end{equation}

\textbf{Corollary 3.}
The time needed for the convergence of the training process is less than the sum of $I_0$ iteration delays in the stochastic order
\begin{equation}
    T_{c}=\sum_{i=0}^{N-1} T^{(i)} \leq_{st} \sum_{i=0}^{I_0-1} T^{(i)}.
\end{equation}
\begin{IEEEproof}
Since $T_{c}=\sum_{i=0}^{N-1} T^{(i)}$  and $N\leq I_0$, we have $ \Pr\{T_c>t\}\leq \Pr\{\sum_{i=0}^{I_0-1} T^{(i)}>t\}$ for all $t>0$.
\end{IEEEproof}

Combined the EVT on one iteration delay with Theorem 4 and Theorem 5, we can get the following asymptotic results for the distribution of overall delay in wireless FL systems when the number of involved users approaches infinity.

\textbf{Theorem 6.} When the number of users in the wireless FL system $K$ approaches infinity, if the conditions in  \eqref{coro1} and \eqref{coro2} are satisfied, then for $\tau> I_0\mathbb{E}[T^{(n)}]$, the tail distribution of $T_c$ can be further expressed as
\vspace{-0.5mm}
\begin{equation}
\Pr\{ T_c \geq \tau \}=\exp(- s^\star \tau)\bigg(\frac{e^{\frac{a}{b}-e^{\frac{a}{b}}}}{b}\int_{0}^{+\infty} e^{(s^\star-\frac{1}{b})y }e^{-e^{-\frac{1}{b}y}} dy\bigg)^{I_0},
\vspace{0.5mm}
\end{equation}
where $a$ and $b$ are chosen differently according to the different downlink schemes from \eqref{coro1a}, \eqref{coro1b} or \eqref{coro2a}, \eqref{coro2b}, while $s^\star$ satisfies \eqref{s*1} or \eqref{s*2} according to the downlink schemes, too.
\begin{IEEEproof}
Please refer to Appendix H for details.
\end{IEEEproof}

It turns out that machine learning models especially deep learning models may not always satisfy the assumptions in Lemma 4. It is thus difficult to find the globally optimal solution for nonconvex machine learning models. In fact, the distributed learning procedures and models used in wireless FL systems become extremely complicated. As mentioned in Section \textrm{I}, the systems and statistical heterogeneity in wireless FL systems brings additional difficulties to perform the exact analysis on $N$. Therefore, by modeling $N$ as a random variable, we can instead use the empirical distribution to characterize it.
 Specifically, we define the probability generating function (PGF) of non-negative discrete random variable of the number of iterations $N$ as 
\begin{equation}
G_N(z)=\sum_{n=0}^{\infty} \Pr\{N=n\}z^n,
\end{equation}
where $\Pr\{N=n\}$ can be approximated by the frequency of $N=n$ in massive independently repeated experiments according to the law of large numbers \cite{ross}. Similarly, the PGF of one iteration delay $T^{(n)}$ is given by
\vspace{-1mm}
\begin{equation}
G_I(z)=\sum_{i=0}^{\infty} \Pr\big\{T^{(n)}=iT_0\big\}z^i,
\vspace{-1mm}
\end{equation}
where $\Pr\big\{T^{(n)}=iT_0\big\}$ can be derived from \eqref{eq25} in the synchronous downlink transmission scheme or \eqref{t2} in the asynchronous downlink transmission scheme. Since the overall delay $T_c$ is a compound random variable of $N$ and $T^{(n)}$, now we can characterize the distribution of $T_c$ based on $G_N(z)$ and $G_I(z)$ in the following corollary.

\textbf{Corollary 4.} Given the PGF $G_N(z)$ of the iteration number for convergence $N$ and $G_I(z)$ of the one iteration delay $T^{(n)}$, the overall delay distribution can be expressed as
\begin{equation}
\Pr\{T_c=dT_0\}=\frac{G^{(d)}_N\big( G_I(z)\big) \big |_{z=0}}{d!},
\vspace{-1mm}
\end{equation}
where $G^{(d)}(\cdot)$ is the $d$-th derivative of $G(\cdot)$.
\begin{IEEEproof}
Please refer to Appendix I for details.
\end{IEEEproof}

\vspace{-3mm}
\section{Theoretical and Simulation results}
In this section, we conduct experiments to validate the accuracy of the theoretical analysis by comparing the theoretical analysis results and empirical simulation results. Based on these results, we provide more insights on the delay distribution in wireless FL systems.
\begin{table}
\small
\centering 
\caption{System parameters}  
\vspace{-8mm}
\begin{center}  
\begin{tabular}{|c|c||c|c|} 
\hline  
\multicolumn{1}{|c|}{\textbf{Parameter}} & \multicolumn{1}{c||}{\textbf{Value}} & \multicolumn{1}{c|}{\textbf{Parameter}} & \multicolumn{1}{c|}{\textbf{Value}} \\ \hline  
$B$ & 100 KHz & $T_0$  & 2.5 ms \\ \hline  
$S_{\rm SVM}$ & 32080 bits  & $S_{\rm CNN}$ & 414160 bits  \\  \hline  
$\lambda$ & 10 dB & $\lambda_d$ & 20 dB \\ \hline
$\mathbb{E}\{|h_{\rm ul}|\}$ & $\sqrt {\pi/2}$ & $\mathbb{E}\{|h_{\rm dl}|\}$ & $\sqrt {\pi/2}$ \\ \hline
$|\mathcal{D}_k|$ & 2000 &   $\xi$ & 0.01   \\ \hline
$\epsilon_0$ & $10^{-3}$ & $\eta$ & $0.01$ \\\hline
\end{tabular}  
\end{center}  
\vspace{-9mm}
\end{table}
\vspace{-1mm}
\subsection{Experiment Settings}

For our simulations, we consider a wireless FL system with $K=30$ users involved to perform an image classification task. We run experiments on the handwritten digit database MNIST \cite{mnist} with support vector machine (SVM) and a  convolutional neural network (CNN) as the classification models. Model parameters are quantized into bit sequence with 8-bit per parameter for transmission. The other system parameters used in the simulations are listed in Table {\rm I}, where $S_{\rm SVM}$ and $S_{\rm CNN}$ are the quantized  size of the corresponding  model parameters. The CNN model used in our simulations consists of one hidden layer with 20 filters, each with a height and width of 5. Each user has the i.i.d. dataset with equal size, which is sampled uniformly from all training samples. We use MATLAB to simulate the FL algorithm, while we use MATLAB and Mathematica in combination to compute the theoretical results with high computation precision.

\vspace{-1mm}
\subsection{Experiment Results}

\begin{figure}[t]
\centering
\subfigure[SVM model.]{\includegraphics[width=7.5cm]{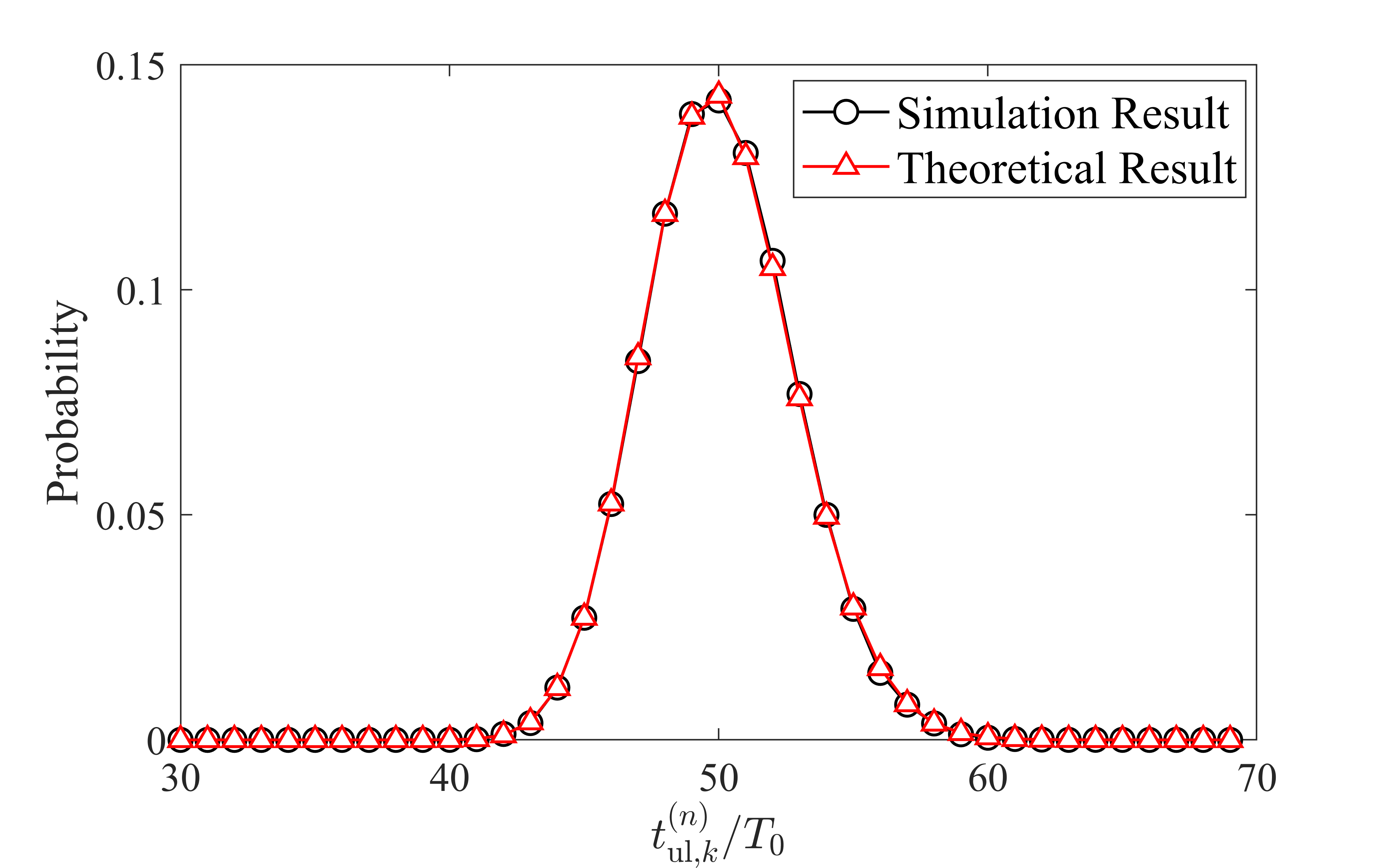}}
\subfigure[CNN model.]{\includegraphics[width=7.5cm]{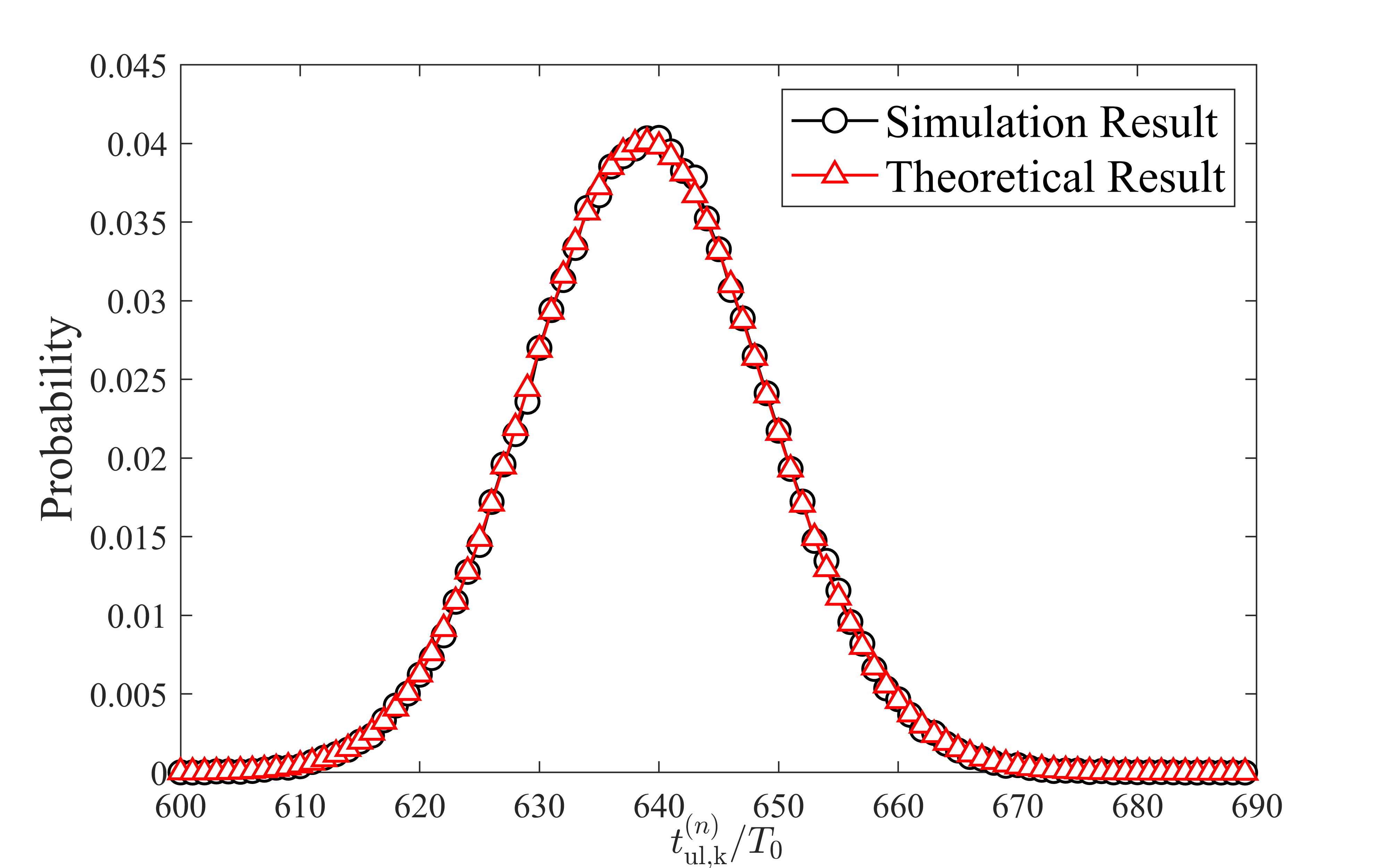}}
\vspace{-3mm}
\caption{Theoretical and simulation results of the one user's uplink delay with different models.}
\label{fig4}
\vspace{-6mm}
\end{figure}

Fig. \ref{fig4} demonstrates the theoretical and simulation results of the one user's uplink delay with different models. Note that, theoretically, the uplink delay can be infinity, but in practical computation and simulation, we neglect the conditions that the uplink delay becomes too large with almost zero probability. From Fig. \ref{fig4}, we observe that the theoretical results obtained via  \eqref{eq20} are almost identical with their corresponding simulation results. The maximal error between numerical results and simulation results is about 3\%. Moreover, since the CNN model has more uploading information, it is reasonable that $\mathbb{E}\{t_{{\rm ul},k}^{(n)}\}$ of the CNN model is much larger than that of the SVM model. 
\begin{figure}[t]
\centering
\subfigure[SVM model.]{\includegraphics[width=7.5cm]{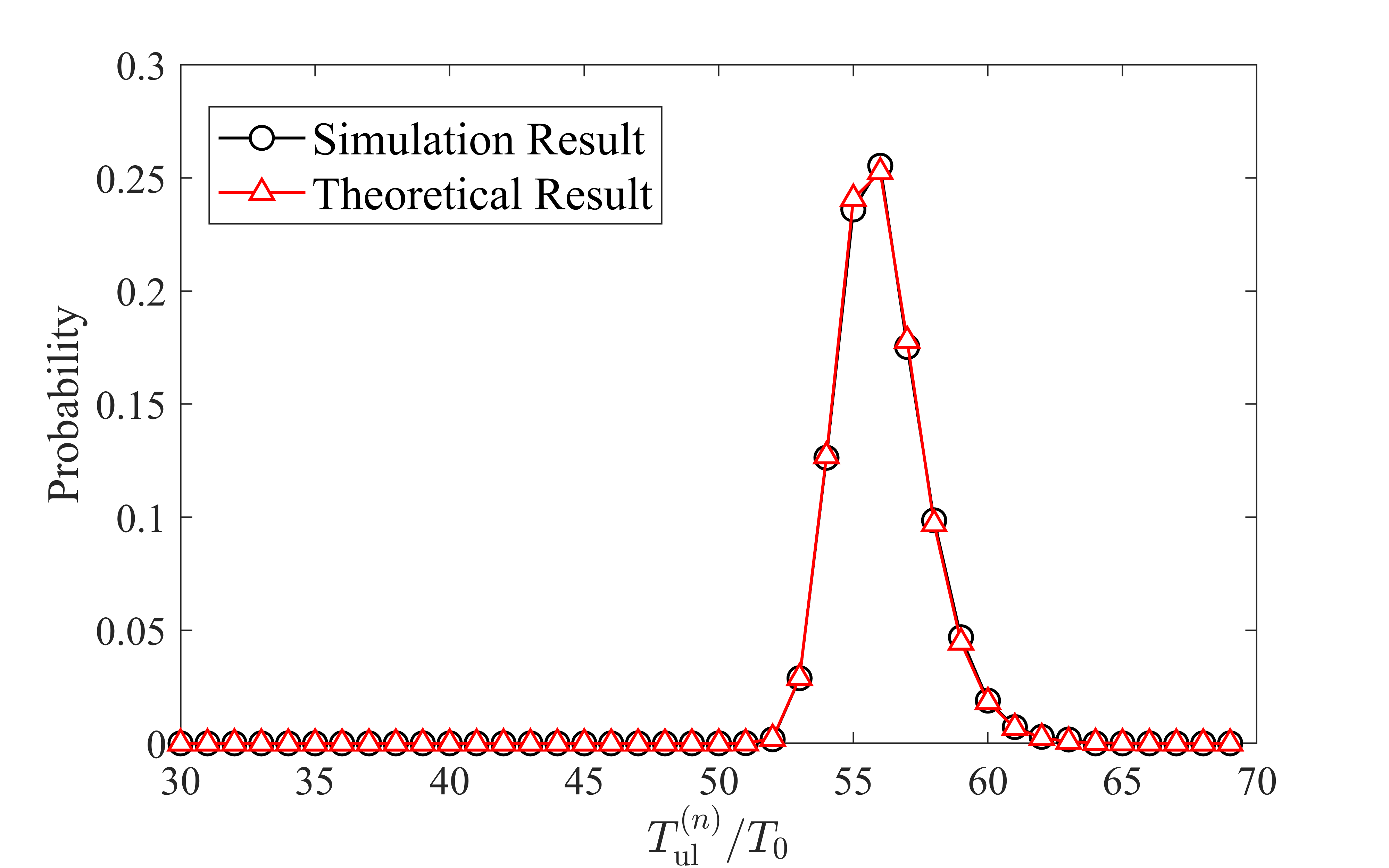}}
\subfigure[CNN model.]{\includegraphics[width=7.5cm]{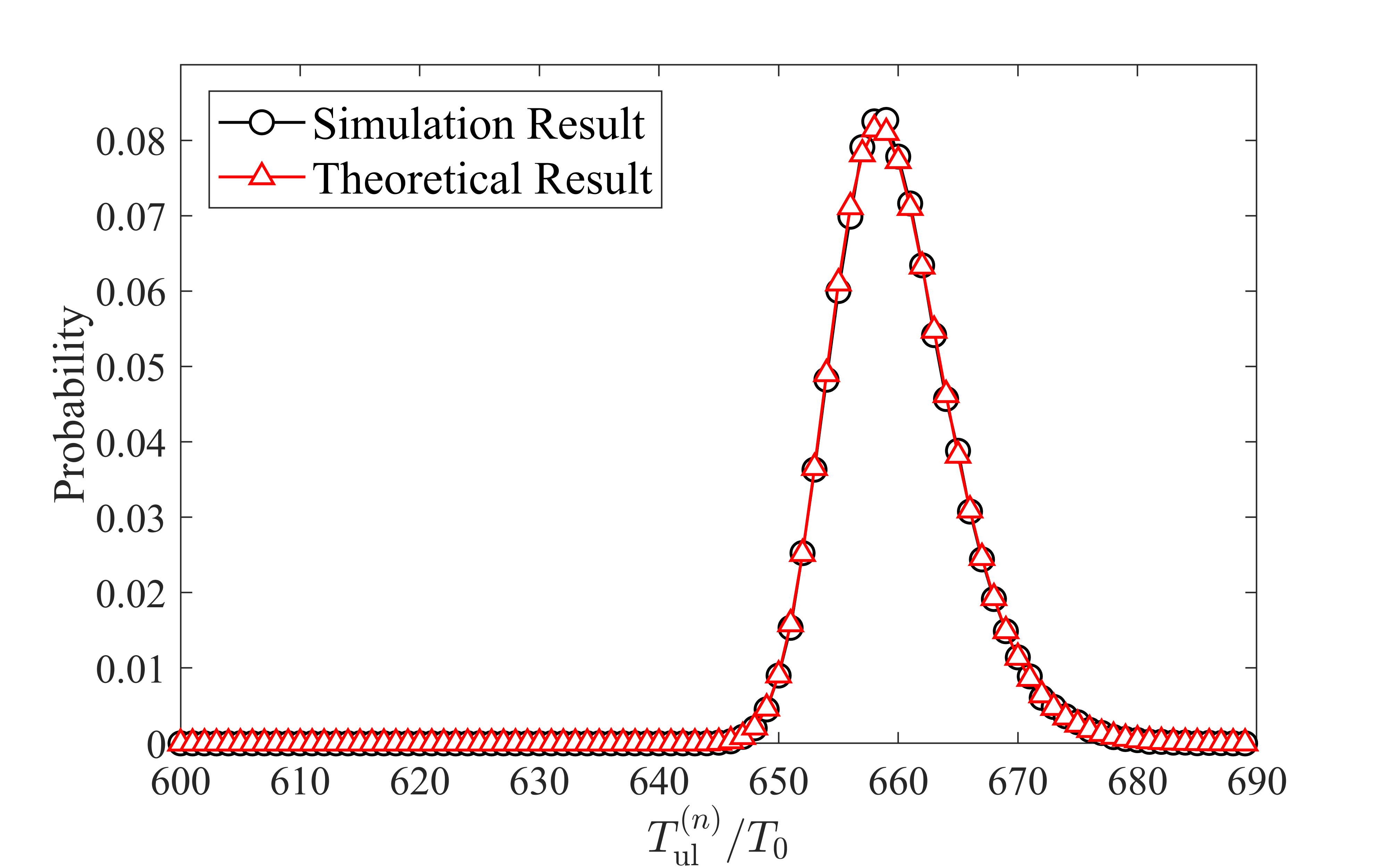}}
\vspace{-2mm}
\caption{Theoretical and simulation results of the one iteration uplink delay in the synchronous downlink scheme.}
\label{fig5}
\vspace{-6mm}
\end{figure}
Furthermore, we compare the empirical distribution of the uplink delay in the synchronous downlink scheme obtained from simulations with the theoretical results computed based on Theorem 1. Experiment results in Fig. \ref{fig5} show that the theoretical results match well with their corresponding simulation results. From these results, we see that the saddle point approximation based method satisfactorily approximates the distribution of $t_{{\rm ul},k}^{(n)}$ and $T_{\rm ul}^{(n)}$ in wireless FL systems even when the number of involved time slots is not relatively large, which shows the effectiveness and advantage of the saddle point approximation method in this problem.

\begin{figure}[t]
\centering
\subfigure[SVM model.]{\includegraphics[width=7.5cm]{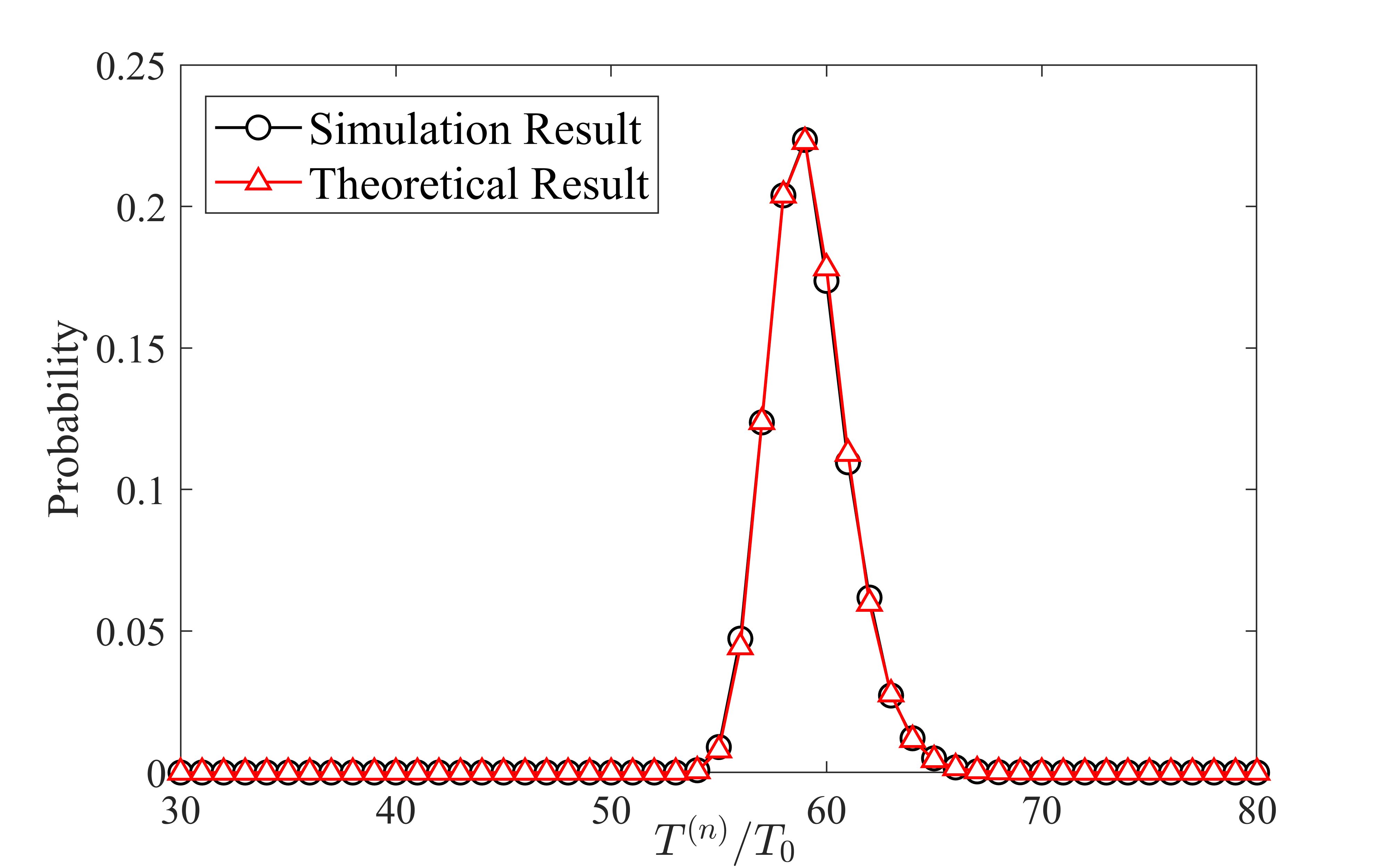}}
\subfigure[CNN model.]{\includegraphics[width=7.5cm]{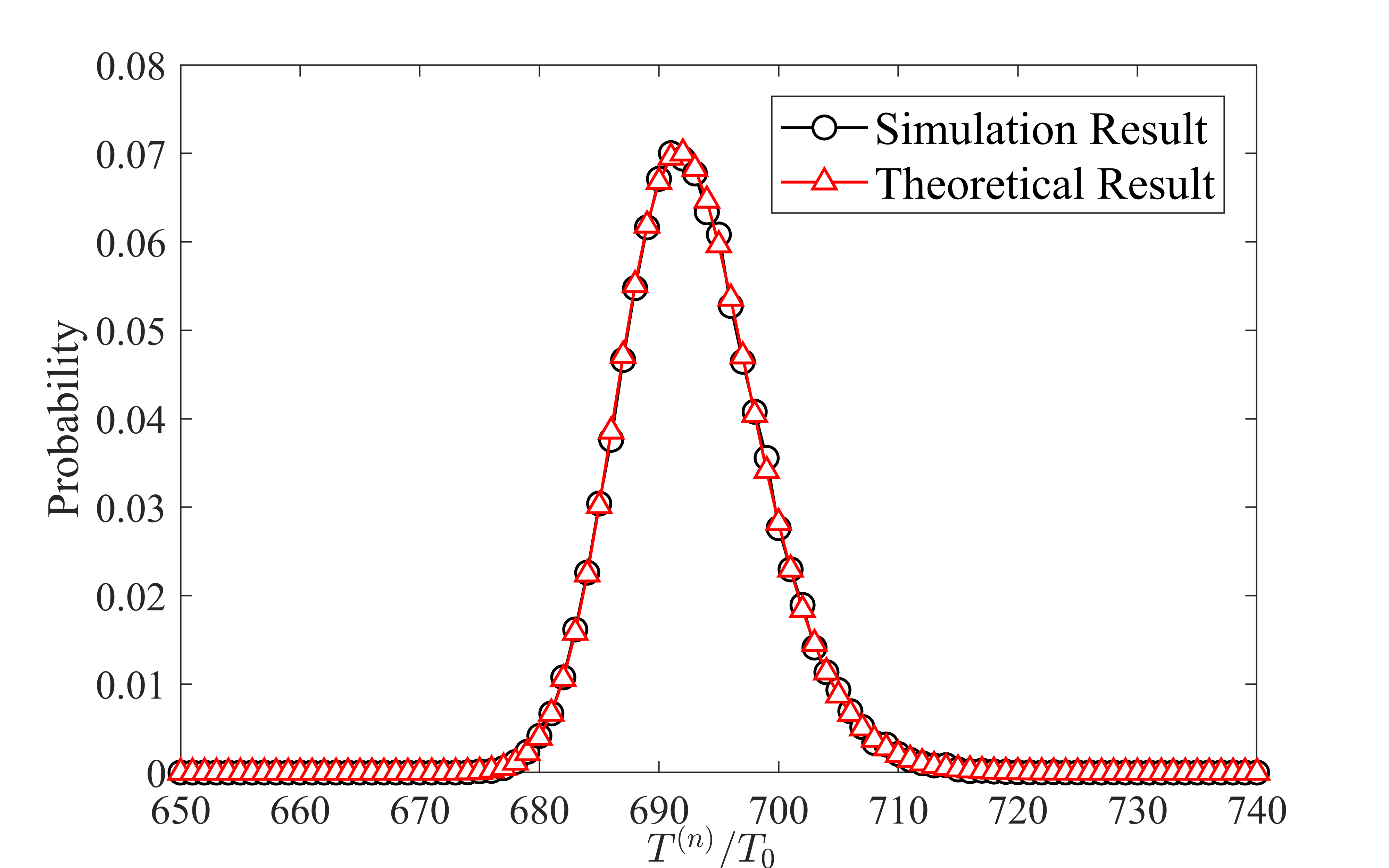}}
\vspace{-2mm}
\caption{Theoretical and simulation results of the one iteration delay in the synchronous transmission scheme.}
\label{fig6}
\vspace{-6mm}
\end{figure}
\begin{figure}[t]
\centering
\subfigure[SVM model.]{\includegraphics[width=7.5cm]{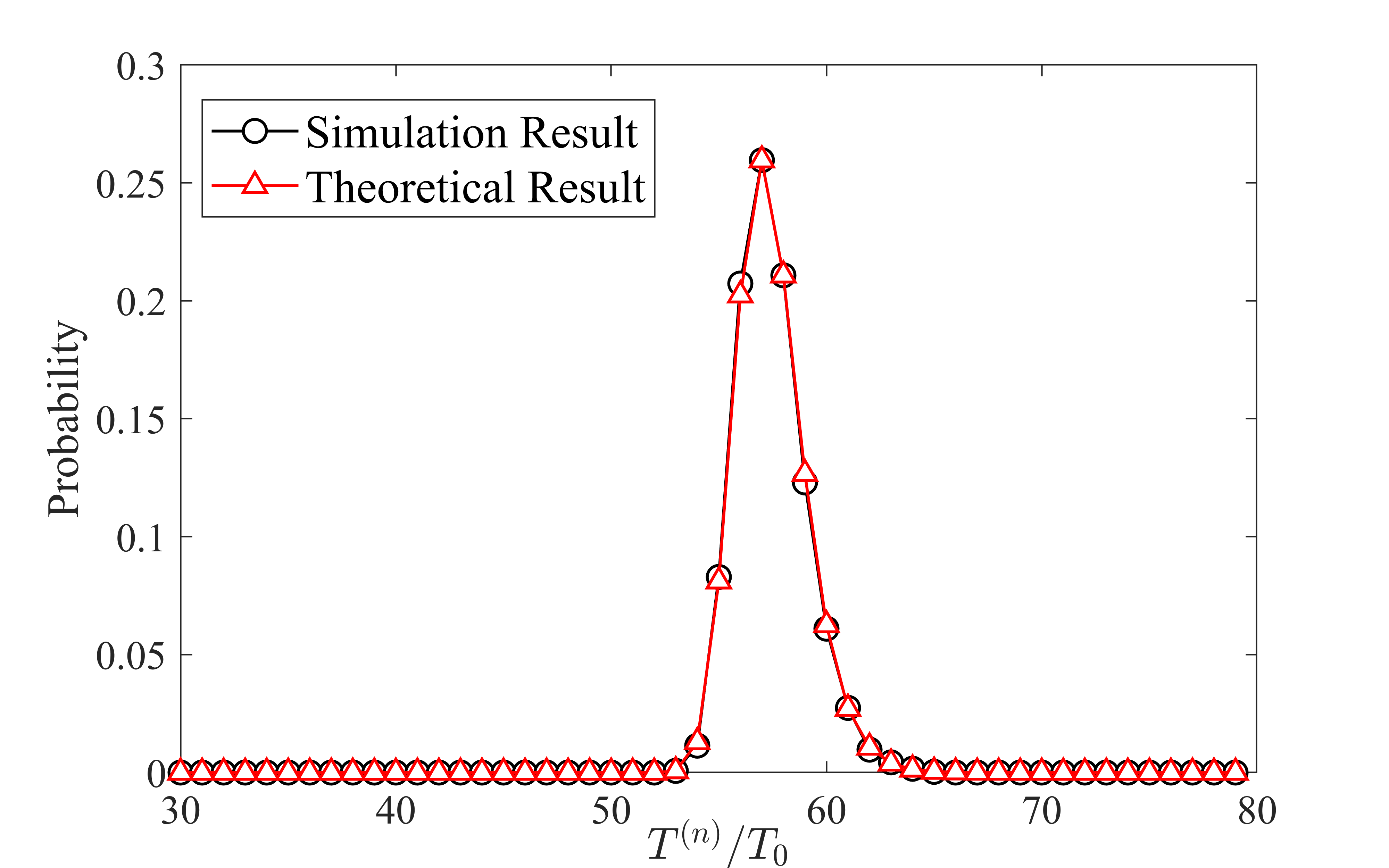}}
\subfigure[CNN model.]{\includegraphics[width=7.5cm]{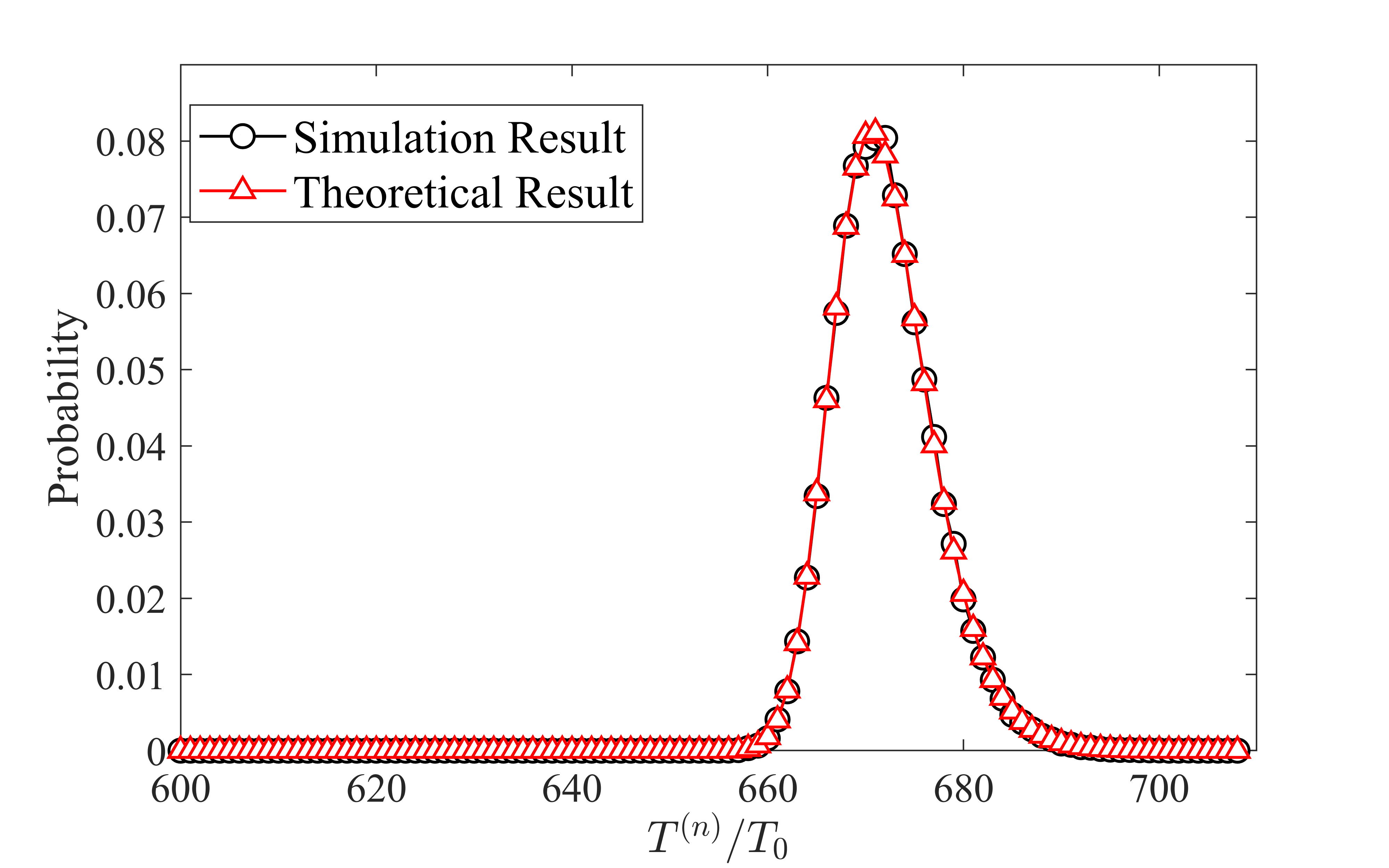}}
\vspace{-2mm}
\caption{Theoretical and simulation results of the one iteration delay in the asynchronous transmission scheme.}
\label{fig7}
\vspace{-6.9mm}
\end{figure}
Fig. \ref{fig6} shows the theoretical and simulation results of one iteration delay in the synchronous downlink scheme with different models. The theoretical results perfectly match their corresponding results, which further validate our theoretical analysis. Besides, by comparing Fig. \ref{fig5} and \ref{fig6}, the corresponding theoretical $\mathbb{E}\{T_{\rm ul}^{(n)}\}$ of SVM and CNN are 0.1351s and 1.6494s, while the corresponding theoretical $\mathbb{E}\{T^{(n)}\}$ of SVM and CNN are 0.1479s and 1.7321s. From this, we find that the one iteration delay is dominated by the uplink delay. This is reasonable since the BS occupies more bandwidth and larger power, which results in the relatively small downlink transmission delay. 
Fig. \ref{fig7} presents the theoretical and simulation results of one iteration delay for the asynchronous downlink scheme with different models. The corresponding theoretical $\mathbb{E}\{T^{(n)}\}$ of SVM and CNN in the asynchronous downlink scheme are 0.1436s and 1.6800s. Comparing these results with that of the synchronous downlink scheme, we find that the asynchronous downlink scheme has better performance on one iteration delay than the synchronous downlink scheme. This is reasonable as the asynchronous downlink scheme fully uses each user's channel gain instead of the worst one. However, since the channel coefficient is i.i.d. over the time slots, the uplink delay of the asynchronous downlink scheme will exhibit no difference with that of the synchronous downlink scheme in terms of statistical results. Meanwhile, the downlink delay is relatively small as we analyzed above. Hence, the performance gain on reducing one iteration delay in the asynchronous downlink scheme is limited. Moreover, the time consumption of one iteration computation $t^{(n)}_{{\rm cp},k}$ in the simulations is about $2\times 10^{-4}$s for SVM and 0.1241s for CNN. Note that, the simulations are conducted on a CPU environment, which means $t^{(n)}_{{\rm cp},k}$ can be further reduced by using more powerful computation processors or adopting training acceleration methods. Therefore, it is reasonable to make the assumption that the computation time consumption can be ignored when we aim to characterize the delay distributions in wireless FL systems.

\begin{figure}[t]
\begin{minipage}[t]{0.5\linewidth}
\centering
\includegraphics[width=7.5cm]{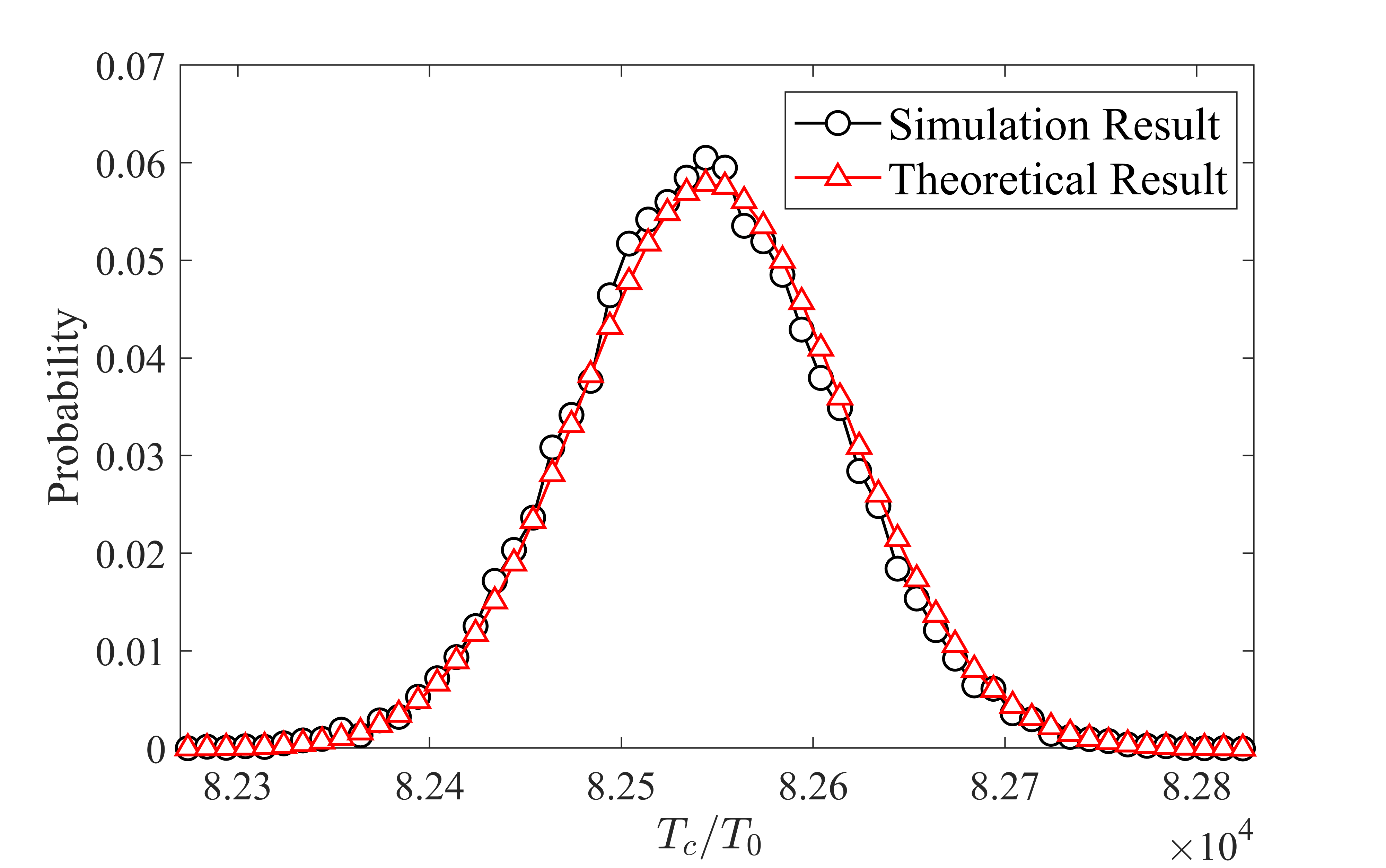}
\vspace{-3mm}
\caption{Distribution of the overall delay under $I_0$ global \quad \\ iterations in the synchronous downlink scheme.}
\centering
\vspace{-6.5mm}
\label{I_0}
\end{minipage}%
\begin{minipage}[t]{0.5\linewidth}
\centering
\includegraphics[width=7.5cm]{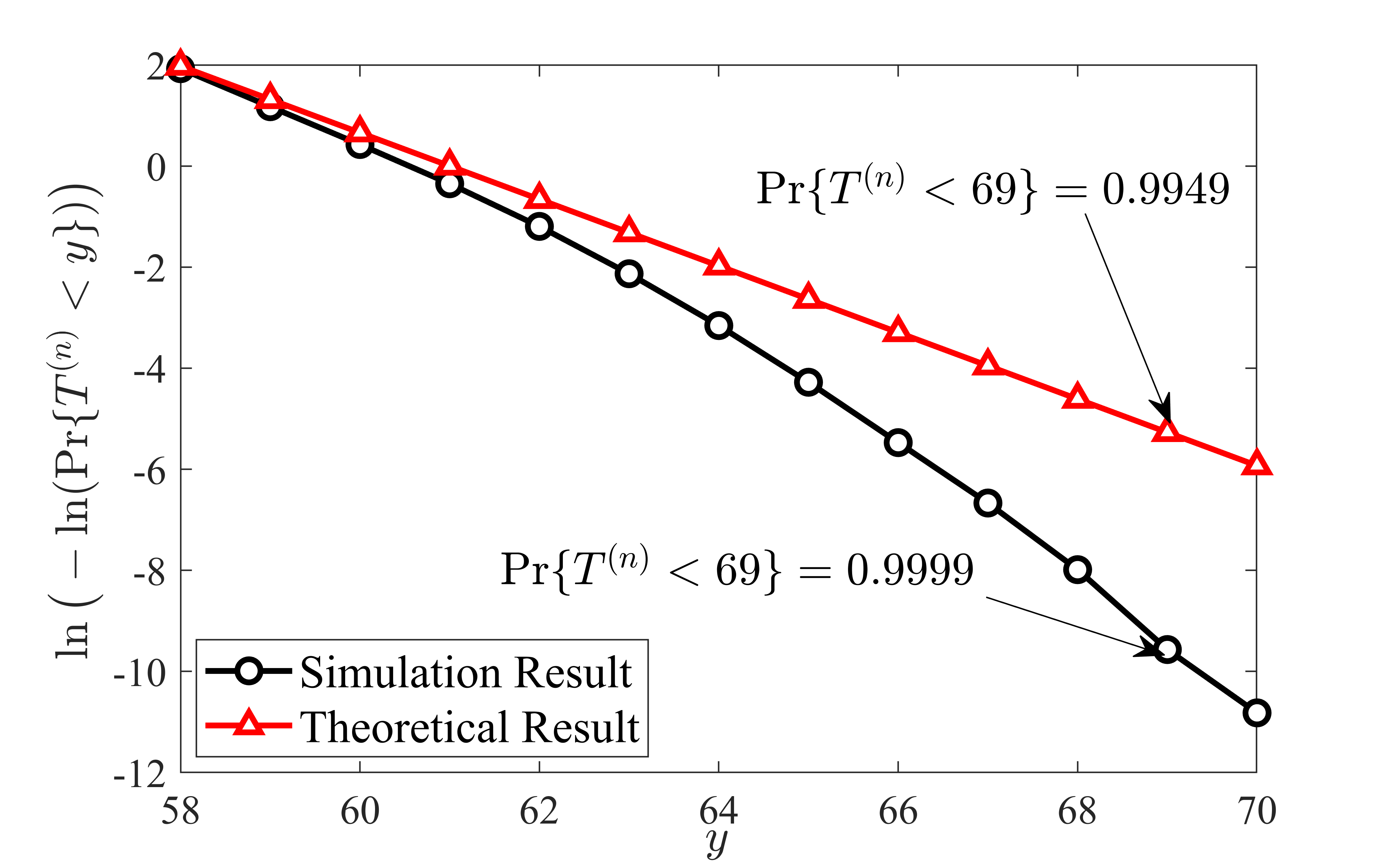}
\vspace{-3mm}
\caption{EVT analysis on one iteration delay when $K=1000$.}
\vspace{-6.5mm}
\label{eva}
\end{minipage}
\end{figure} 

\begin{figure}[t]
\begin{minipage}[t]{0.5\linewidth}
\centering
\includegraphics[width=7.5cm]{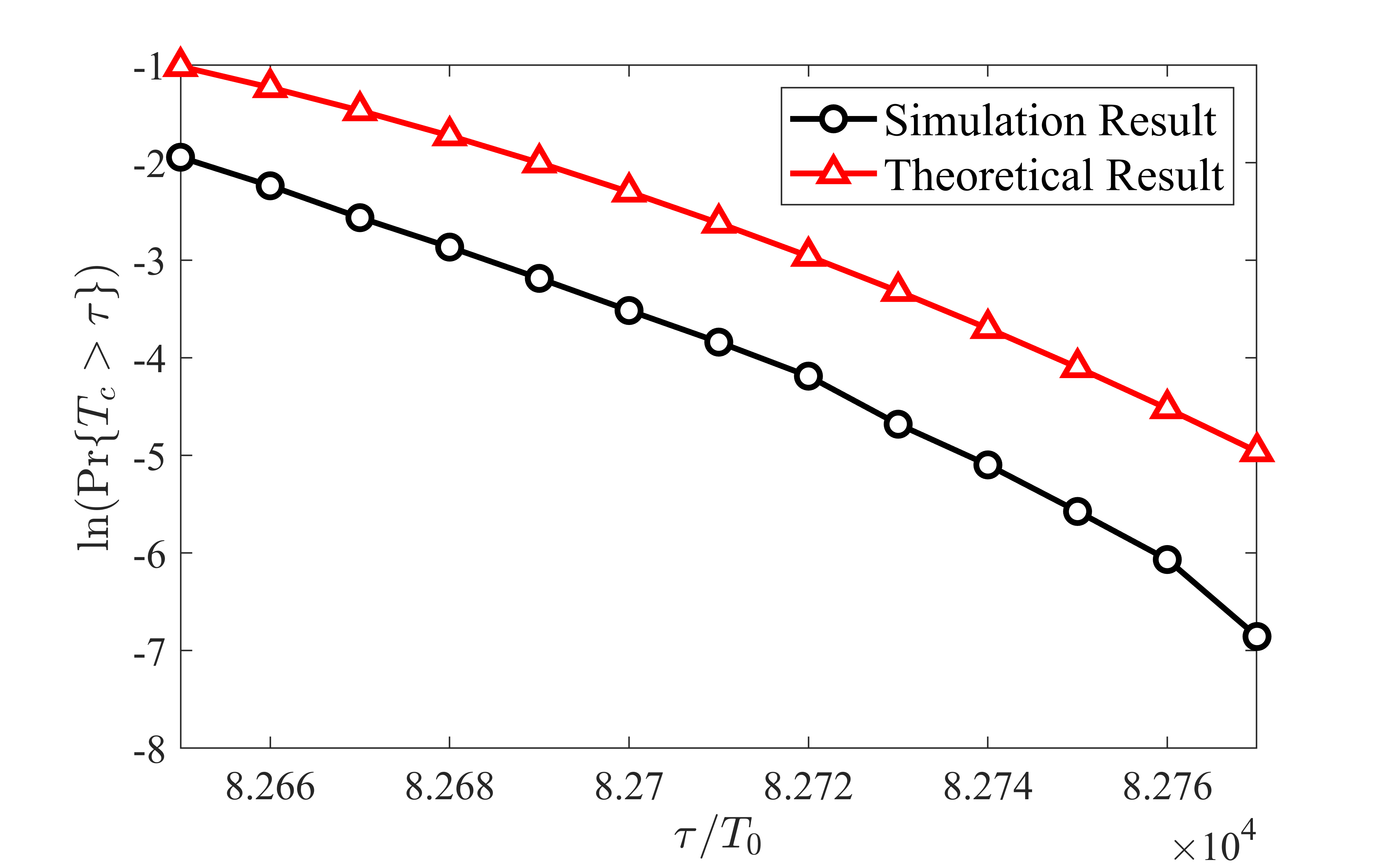}
\vspace{-3mm}
\caption{LDT analysis on the distribution of the overall \quad \quad \\ delay.}
\centering
\vspace{-6mm}
\label{ldt}
\end{minipage}%
\begin{minipage}[t]{0.5\linewidth}
\centering
\includegraphics[width=6cm]{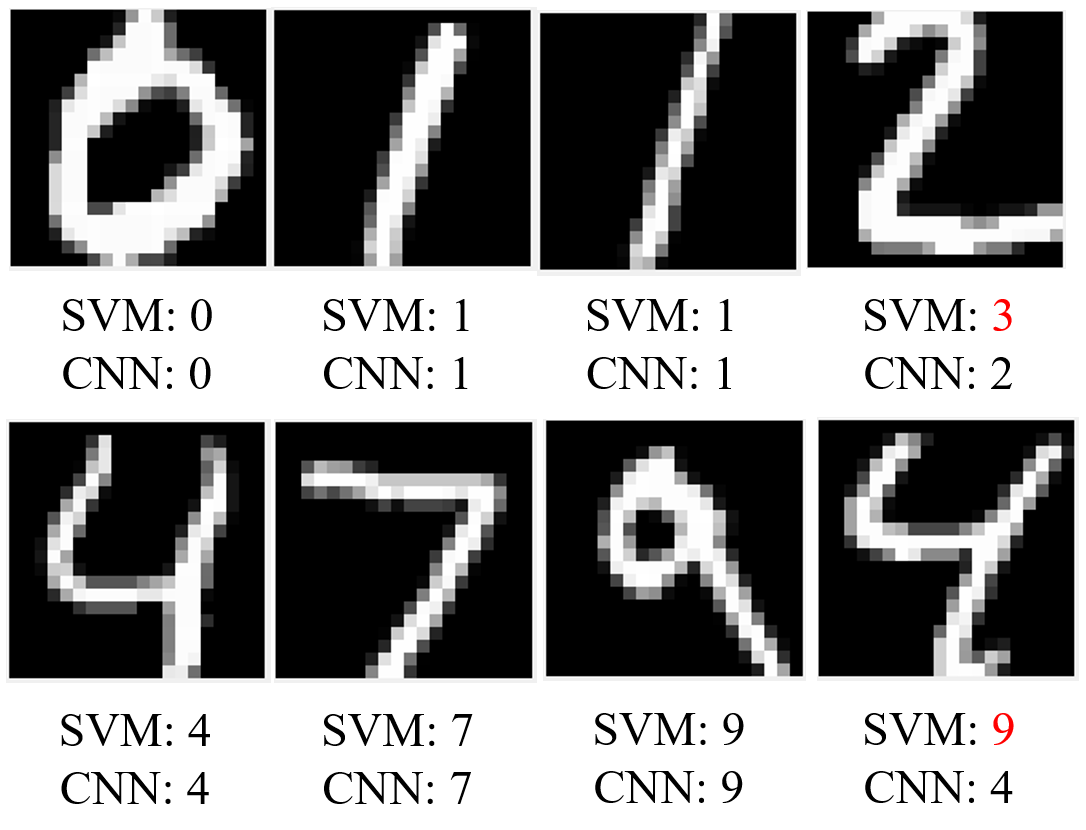}
\vspace{-3mm}
\caption{Examples of implementing FL for handwritten digit image identification.}
\vspace{-6mm}
\label{figres}
\end{minipage}
\end{figure}

\begin{figure}[t]
\centering
\subfigure[SVM model.]{\includegraphics[width=7.5cm]{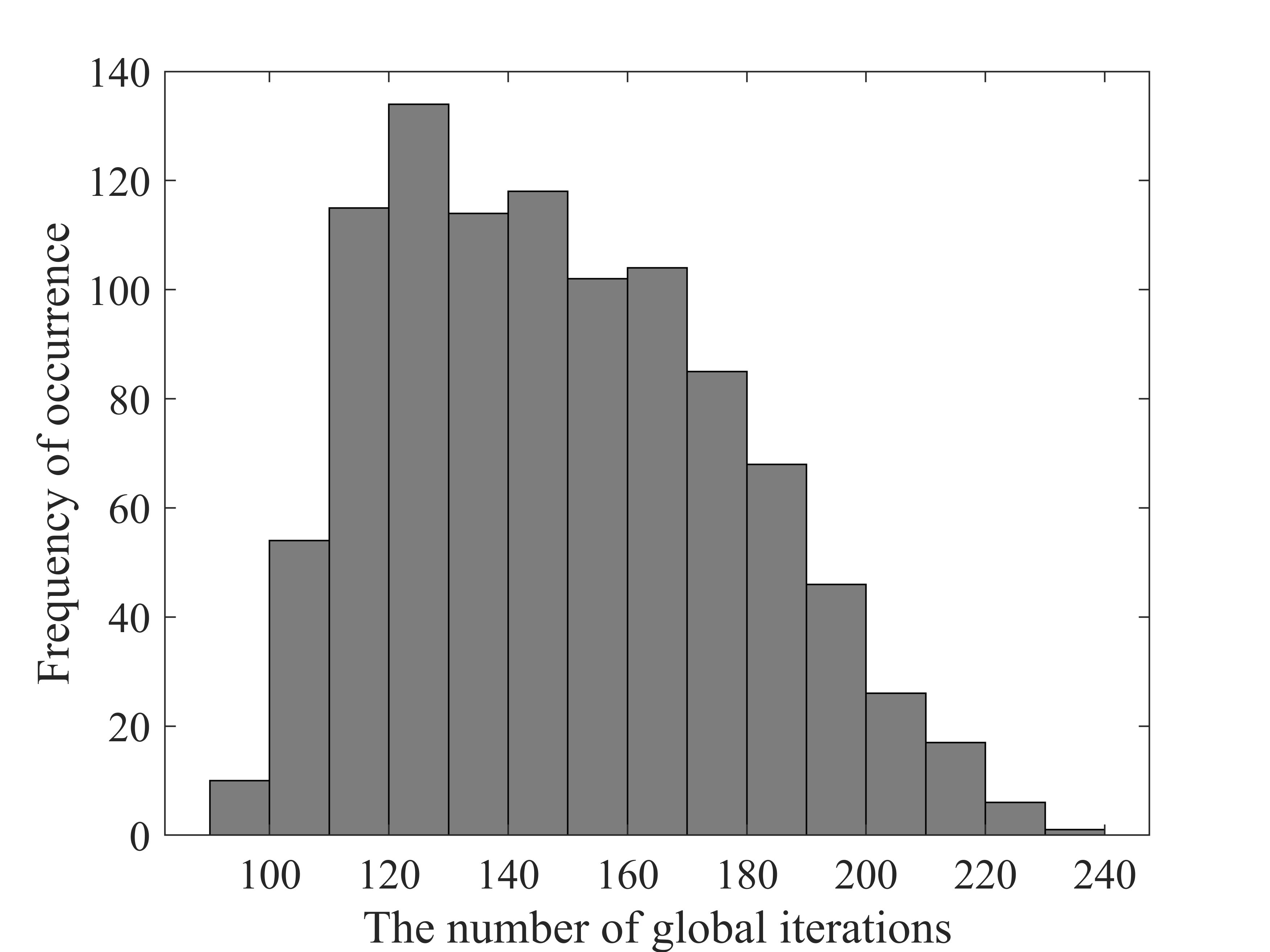}}
\subfigure[CNN model.]{\includegraphics[width=7.5cm]{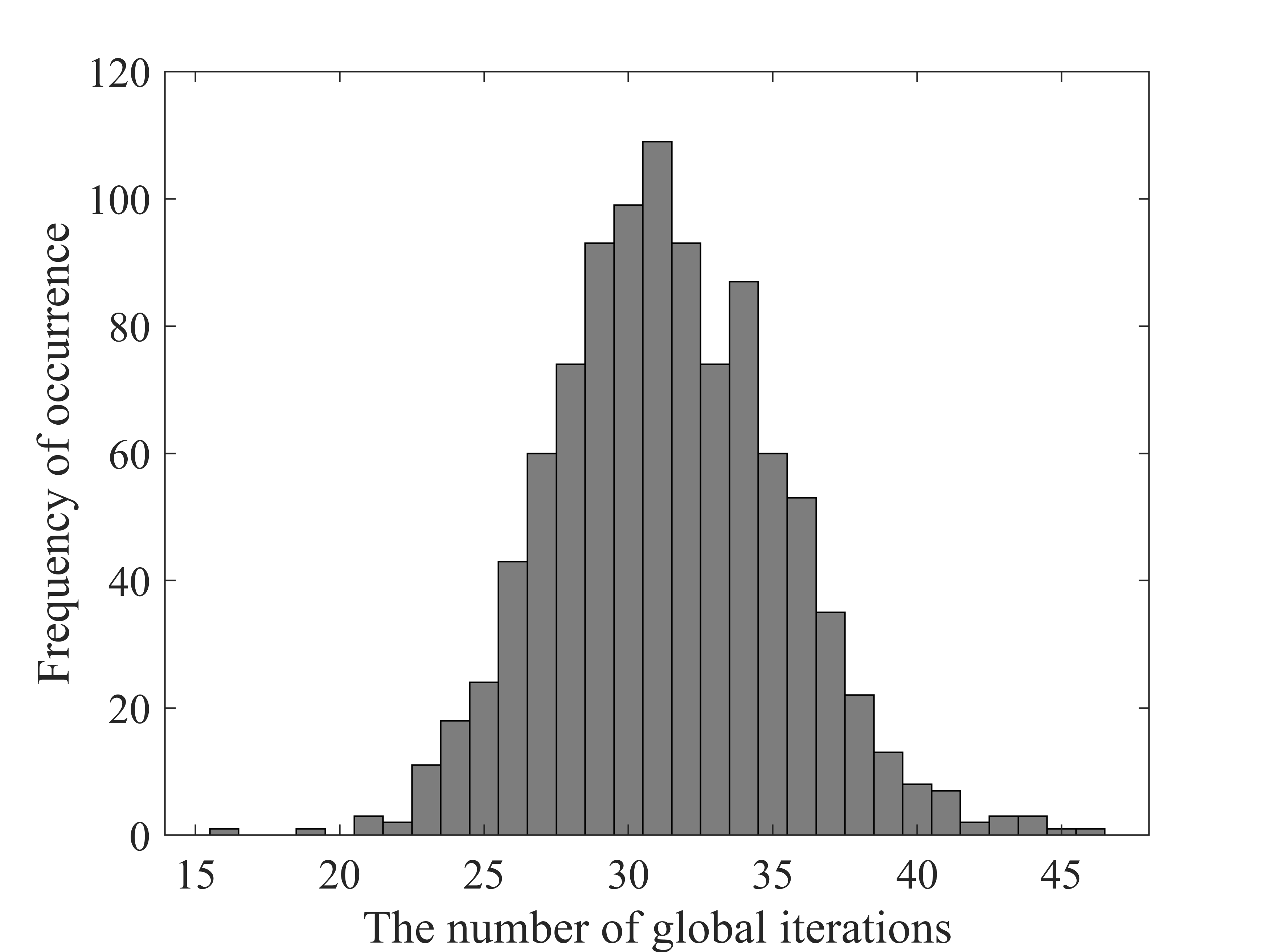}}
\vspace{-1mm}
\caption{The frequency distribution histogram of iteration numbers for convergence.} \label{fig8}
\vspace{-6mm}
\end{figure}
\begin{figure}[t]
\centering
\subfigure[SVM model.]{\includegraphics[width=7.5cm]{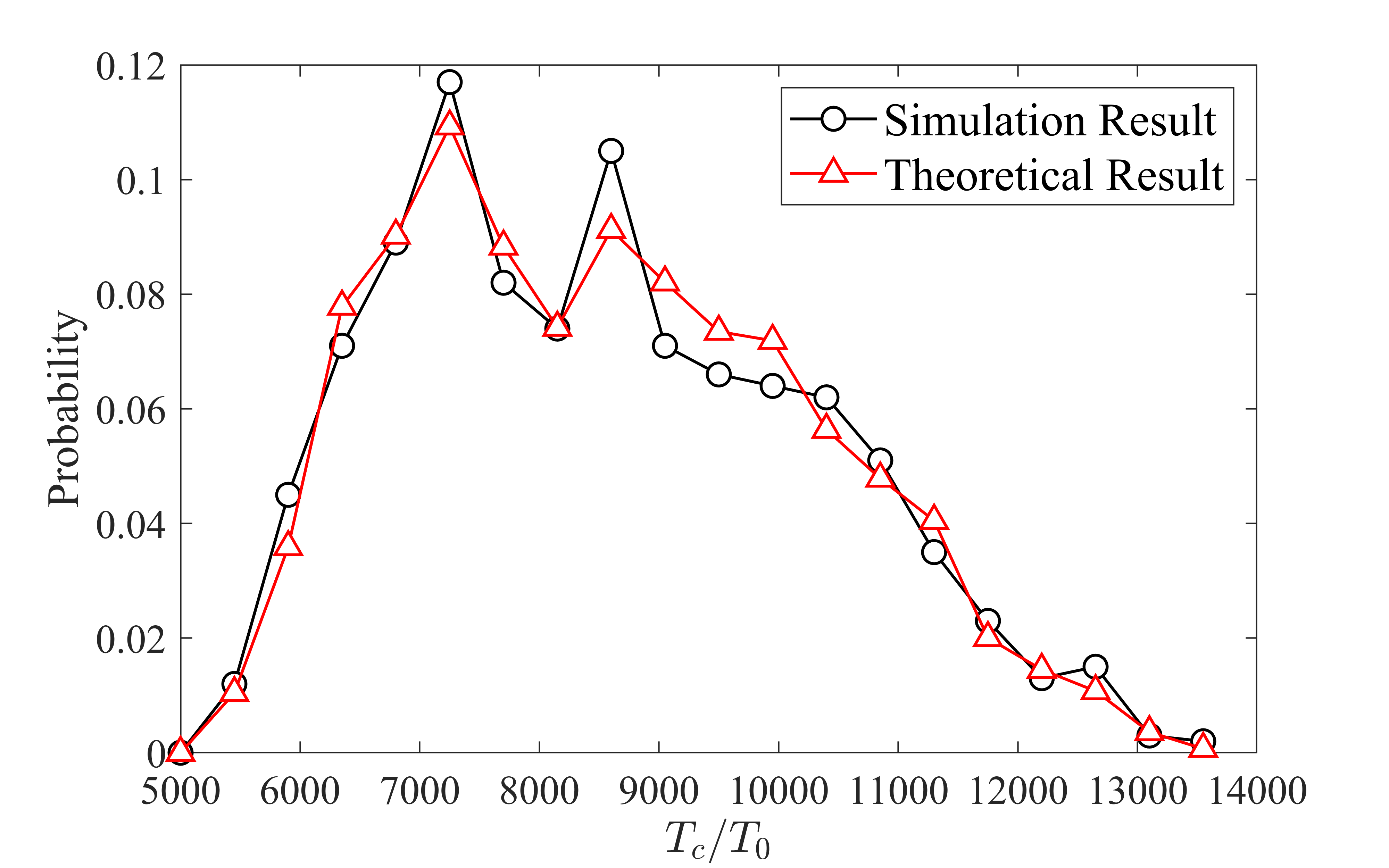}}
\subfigure[CNN model.]{\includegraphics[width=7.5cm]{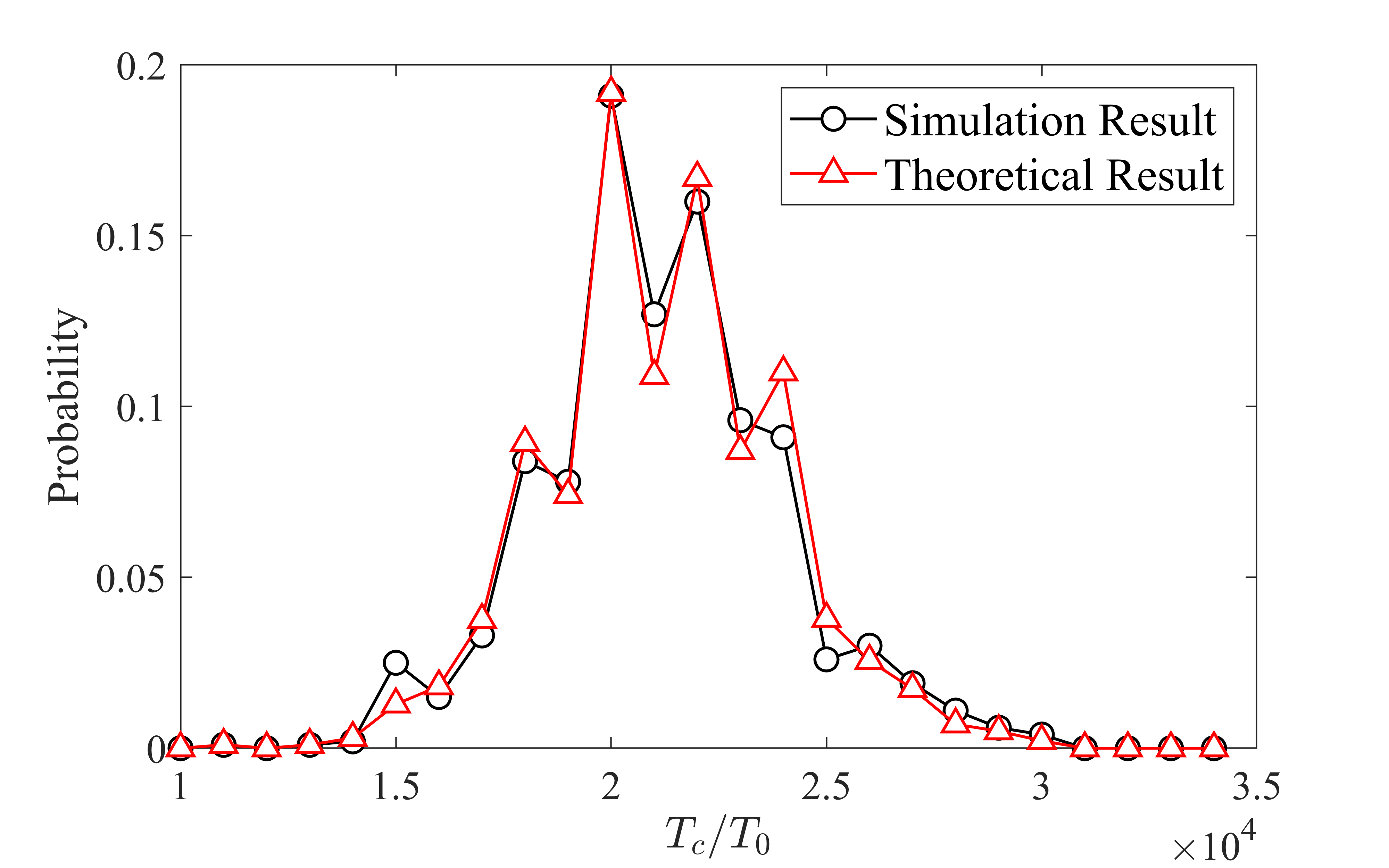}}
\vspace{-1mm}
\caption{The overall delay distribution in the synchronous downlink scheme.} \label{fig9}
\vspace{-5mm}
\end{figure}
Fig. \ref{I_0} shows the theoretical and simulation results of the overall delay distribution with the SVM model and the synchronous downlink scheme under $I_0$ iterations. For the SVM model we used in this experiment, $\gamma / \alpha$ is 1. Recalling Lemma 4 and the parameters shown in Table \textrm{I}, $I_0$ is equal to 1395 in this experiment. Note that, since one iteration delay is a discrete variable, once we get the accurate distribution of one iteration delay, the convolution result computed by software is precise. However, the approximation errors of one iteration delay distribution are accumulated over $I_0$ times. Thus, the numerical results are not perfectly identical with the simulation results like Fig. \ref{fig6}(a), which is actually reasonable.    

To provide more insights, we focus on the asymptotic properties in this wireless FL system. We conduct the experiments on the SVM model by adopting the synchronous downlink scheme. Fig. \ref{eva} shows the theoretical and simulation results of the one iteration delay when $K=1000$. From this figure, we can see that after taking the logarithm twice, the simulation result shows the linear behavior of variable $y$ and the intercepts of the two curves is close, which validates Corollary 1. However, the slope of the simulation curve seems to become smaller with $y$ increasing. The reason is that the number of simulations is not probably large enough to reflect the true distribution of $T^{(n)}$, while the twice logarithm operation magnifies the difference between them.  
Fig. \ref{ldt} represents the theoretical and simulation results of the overall delay under $I_0$ iterations. By taking the logarithm, the slope of simulation results is approximately identical with the theoretical one, which validates Theorem 4. When $\tau$ is large, the slope becomes smaller, which means $\Pr \{T_c \geq \tau\}$ decays faster than the exponential distributions.  

Fig. \ref{figres} persents performance results of the proposed FL algorithm for handwritten digit identification. Fig. \ref{fig8} is the frequency distribution histogram of the number of iterations for convergence. In Fig. \ref{fig8}(a), we can see that all numbers are less than a threshold in this simulation for SVM, which proves that Lemma 4 has some significance. Moreover, from Fig. \ref{fig8}(b), we can see that although CNN does not meet the assumption mentioned in Lemma 4, it shows the same property with the result of SVM. However, since CNN is a nonconvex model, it may not converge. Although probability is rather small to be ignored in the simulation, $N$ can be infinity in theory, which means there may not be an upper bound for nonconvex models. Hence, by taking all of the above factors into account, it is reasonable to treat $N$ as a random variable.  

Fig. \ref{fig9} represents both the empirical and theoretical distribution of the overall delay in the synchronous downlink scheme with different models. The difference between the empirical result and the theoretical result is slightly more evident compared to Fig. 6. It is due to the fact that we use the empirical distribution of $N$ obtained in Fig. \ref{fig8} to get the overall delay distribution, which magnifies the errors. However, the results in Fig. \ref{fig9} still show that the empirical distributions of the overall delay are in good agreement with the theoretical distributions. From this experiment, we see that it is practical to calculate the generating function of $N$ through its empirical distribution. Nonetheless, it is clear that getting more accurate distribution of global iteration numbers remains an open and challenging problem.

\section{Conclusions}
In this paper, we investigated the delay distribution for wireless FL systems. 

In particular, synchronous and asynchronous downlink transmission schemes were further proposed in wireless FL systems, followed by the definitions of the one iteration and overall delay. To characterize the accurate approximation and asymptotic properties of delay distributions in wireless FL systems, saddle point approximation, EVT and LDT methods were exploited. Moreover, we proposed to model the convergence rounds as a random variable instead of a constant, yielding a novel way to provide an empirical distribution for the overall delay in wireless FL systems.
The proposed modeling and analytical tools for the delay distribution provide a novel research direction to better understand and design wireless FL systems in both practical and theoretical aspects. Future promising research directions include improving the resource efficiency and users' experience by leveraging the presented analytical results in this paper.

\appendices
\section{Proof of Lemma 1}
\vspace{-1mm}
According to \eqref{eq9}, we have
\begin{equation}
    t_{{\rm ul},k}^{(n)}=T_0\left\{d: \sum_{t=1}^{d} r_{{\rm ul},k}^{(n)}(t) \geq \frac{S}{T_0},\, \sum_{t=1}^{d-1} r_{{\rm ul},k}^{(n)}(t) < \frac{S}{T_0},\, d \in \mathbb{N}_+\right\},\label{ap1}
\end{equation}
\begin{equation}
    \Pr\big\{ t_{{\rm ul},k}^{(n)} > dT_0\big\}=\Pr\left\{\sum_{t=1}^{d} r_{{\rm ul},k}^{(n)}(t) < \frac{S}{T_0}\right\}.\label{ap2}
\vspace{-1mm}
\end{equation}

Since $r_{{\rm ul},k}^{(n)}(t)$ derives from \eqref{eq13}, where $h_{{\rm ul},k}^{(n)}(t)$ is i.i.d. over $n$ and $t$, for simplicity, we omit $n$ and $t$ in the following derivation in this paper. According to the definition of $Z_d$ in \eqref{zd}, \eqref{ap1} and \eqref{ap2} can be rewritten as 
\vspace{-1mm}
\begin{equation}
t_{{\rm ul},k}=\{dT_0: Z_d\leq 0,\, Z_{d-1}>0,\, d \in \mathbb{N}_+\}.\label{ap3}
\end{equation}
\begin{equation}
    \Pr\{ t_{{\rm ul},k}> dT_0\}=\Pr\{Z_d>0\}.\label{ap4}
\vspace{-5mm}
\end{equation}

Then we obtain 
\vspace{-2mm}
\begin{align}
    \Pr\{t_{{\rm ul},k}=dT_0\}&=\Pr\{ t_{{\rm ul},k}> (d-1)T_0\}-P\{ t_{{\rm ul},k}> dT_0\} \nonumber\\
    &=\Pr\{Z_{d-1}>0\}-\Pr\{Z_d>0\}.\label{ap5}
\end{align}

\vspace{-2mm}
From \eqref{ap5}, we see that if we want to obtain the distribution of $t_{{\rm ul},k}$, we need to get the complementary cumulative distribution function (CCDF) of $Z_d$ first. However, we can not get the distribution of $Z_d$ easily, since it is hard to obtain the distribution of $\sum_{t=1}^{d} r_{{\rm ul},k}$ by convolution. Therefore, we turn to saddle point approximation to obtain an accurate approximation of $Z_d$'s CCDF.
 
CGF is defined as the natural logarithm of moment generating function (MGF). Hence, the CGF of $Z_d$ is given by
\vspace{-4mm}
\begin{align}
    K_d(s)&=\ln{\mathbb{E}\big\{e^{sZ_d}\big\}}\nonumber\\
    &=\frac{S}{BT_0}s+\ln{\mathbb{E}\big \{e^{- s \frac{\sum_{t=1}^{d} r_{{\rm ul},k}}{B}}\big \}}\nonumber\\
    &=\frac{S}{BT_0}s+d\ln{\mathbb{E}\big \{e^{- \frac{sr_{{\rm ul},k}}{B}}\big \}}.\label{ap6}
\end{align}
Similar to MGF, CGF allows us to obtain the $n$-th cumulant of $Z_d$ by evaluating its $n$-th derivative at zero. The first two cumulants are the mean and the variance. 

Having the CGF of $Z_d$, we need to find the saddle point $s^*_d$. Since we want to obtain $\Pr\{Z_d > 0\}$, we have to solve the following saddle point equation:
\vspace{-1mm}
\begin{equation}
K_d^{'}\big(s^*_d\big)=0.\label{ap6-7}
\vspace{-2mm}
\end{equation}
Then the CCDF of $Z_d$ can be approximated by the first two terms of the LR formula as
\begin{equation}
    \Pr\{ Z_d>0\} = 1-\frac{1}{\sqrt{2\pi}}\int_{-\infty}^{\omega_d}e^{-\frac{u^2}{2}}du+\frac{e^{-\frac{\omega_d^2}{2}}}{\sqrt{2\pi}}\Big(\frac{1}{\psi_d}-\frac{1}{\omega_d}\Big).\label{LRappa}
\vspace{1mm}
\end{equation}
The $\omega_d$ and $\psi_d$, for $d\geq1$, are defined as
\vspace{-1mm}
\begin{equation}
\omega_d={\rm sign}\big(s^*_d\big)\sqrt{-2K_d \big(s^*_d \big)},
\end{equation}
\begin{equation}
\psi_d=s^*_d\sqrt{K_d''\big(s^*_d \big)},
\vspace{-3mm}
\end{equation}
where ${\rm sign}( \cdot )$ is the sign function. When $d=0$, $\Pr\{Z_0>0\}=1$. To keep the definition rigorous, we let $\omega_0=\psi_0=-\infty$.

Given that $|h_{{\rm ul},k}|$ follows Rayleigh distribution $f_h(x)=xe^{-\frac{x^2}{2}}$, we can express the distribution of $\frac{r_{{\rm ul},k}}{B}$ as
\vspace{-1mm}
\begin{equation}
    f_{\rm ul}(r)=\frac{1}{2\lambda}e^{\frac{1}{2\lambda}}e^{r-\frac{e^r}{2\lambda}}\mathbb{I}\{r \geq 0\},
\label{ap7}
\end{equation}
where $\mathbb{I}\{\cdot\}$ is the indicator function.

According to the definition of MGF, the MGF of $\frac{r_{{\rm ul},k}}{B}$ is given by
\vspace{-1mm}
\begin{align}
    M(s)&=\int_{-\infty}^{+\infty}e^{sr}f_{\rm ul}(r)dr \nonumber\\
    &=e^{\frac{1}{2\lambda}}(2\lambda)^s\Gamma\Big(s+1,\frac{1}{2\lambda}\Big).\label{ap8}
\end{align}

Thus, substituting \eqref{ap8} into \eqref{ap6},  we have
\begin{align}
    K_d(s)&=\frac{S}{BT_0}s+d\ln{M(-s)} \nonumber\\
    &=\frac{S}{BT_0}s+d\ln\left(e^{\frac{1}{2\lambda}}(2\lambda)^{-s}\Gamma\Big(-s+1,\frac{1}{2\lambda}\Big)\right).\label{ap9}
\end{align}

\vspace{1mm}
Then we can obtain the first and second derivative of $K_d(s)$ respectively as
\vspace{1mm}
\begin{equation}
    K_d'(s)=\frac{S}{BT_0}-d\frac{M'(-s)}{M(-s)},\label{ap10}
\vspace{1mm}
\end{equation}
\begin{equation}
    K_d''(s)=d\frac{M''(-s)}{M(-s)}-d\frac{\big(M'(-s)\big)^2}{\big(M(-s)\big)^2}.\label{ap11}
\vspace{-1mm}
\end{equation}
$M'(-s)$ and $M''(-s)$ are given by
\begin{equation}
    M'(-s)=e^{\frac{1}{2\lambda}}(2\lambda)^{-s}\bigg [\ln{(2\lambda)}\Gamma\Big(-s+1,\frac{1}{2\lambda}\Big)+\Gamma'\Big(-s+1,\frac{1}{2\lambda}\Big)\bigg],\label{ap12}
\vspace{1mm}
\end{equation}
\begin{equation}
     M''(-s)\!=\!e^{\frac{1}{2\lambda}}(2\lambda)^{-s}\bigg[\ln^2{(2\lambda)}\Gamma\Big(\!-s  +  1,\frac{1}{2\lambda}\Big) \!+2\ln{(2\lambda)}\Gamma'\Big(\!-s +1,\frac{1}{2\lambda}\Big)+\Gamma''\Big(\!-s  + 1,\frac{1}{2\lambda}\Big)\bigg],\label{ap13}
\end{equation}
where $\Gamma'(\cdot)$ and $\Gamma''(\cdot)$ are given by
\begin{equation}
    \Gamma'\Big(-s+1,\frac{1}{2\lambda}\Big)=\Gamma\Big(-s+1,\frac{1}{2\lambda}\Big)\ln\Big(\frac{1}{2\lambda}\Big)+\frac{1}{2\lambda}G_{2,3}^{3,0}\Bigg(\begin{array}{c}0,0\\-s,-1,-1\end{array}\Bigg\vert \frac{1}{2\lambda}\Bigg),\label{ap14}
\end{equation}

\vspace{-3mm}
\begin{align}
    \Gamma''\Big(-s+1,\frac{1}{2\lambda}\Big)&=\Gamma\Big(-s+1,\frac{1}{2\lambda}\Big)\ln^2\Big(\frac{1}{2\lambda}\Big)+\frac{1}{\lambda}\Bigg[G_{3,4}^{4,0}\Bigg(\begin{matrix}0,0,0\\-s,-1, -1,-1\end{matrix}\Bigg|\frac{1}{2\lambda}\Bigg) \nonumber\\
     &+\ln\Big(\frac{1}{2\lambda}\Big)G_{2,3}^{3,0}\Bigg(\begin{matrix}0,0\\-s,-1,-1\end{matrix}\Bigg|\frac{1}{2\lambda}\Bigg)\Bigg].\label{ap15} 
\end{align}
In \eqref{ap14} and \eqref{ap15}, $G_{\rho_3,\rho_4}^{\rho_1,\rho_2}\Bigg(\begin{matrix}a_1,a_2,\ldots,a_{\rho_3}\\ b_1,b_2,\ldots,b_{\rho_4} \end{matrix}\Bigg|z\Bigg)$ is the Meijer G-function, which is defined in \eqref{gfunc}.

Combined with \eqref{ap6-7}, \eqref{ap10}, \eqref{ap12} and \eqref{ap14}, we can know that the parameter $s^*_d$ satisfies the following equation:
\vspace{-1mm}
\begin{equation}
\frac{S}{BT_0}=\frac{dG_{2,3}^{3,0}\Bigg(\begin{array}{c}0,0\\-s^*_d,-1,-1\end{array}\Bigg\vert \frac{1}{2\lambda}\Bigg)}{2\lambda \Gamma(-s^*_d+1,\frac{1}{2\lambda})}.
\vspace{1mm}
\end{equation}

Once we get $s^*_d$, $\omega_d={\rm sign}\big(s^*_d\big)\sqrt{-2K_d \big (s^*_d\big)}$ can be obtained from  \eqref{ap9}. Moreover, $\psi_d=s^*_d\sqrt{K_d''(s^*_d)}$ can also be obtained by using \eqref{ap11}, \eqref{ap12}, \eqref{ap13}, \eqref{ap14}, and \eqref{ap15}. Hence, according to the LR formula, we can approximate the probability $\Pr\{Z_d>0\}$ by \eqref{LRappa}. 
Substituting \eqref{LRappa} into \eqref{ap5}, the distribution of one user's uplink time slots consumption is obtained as \eqref{eq20} shows. 

\vspace{-3mm}
\section{Proof of Theorem 1}
As shown in \eqref{eq15}, we notice that $T_{\rm ul}$ is the maximum order statistic between one user's uplink delay $t_{{\rm ul},k}$. Therefore, the probability of $T_{\rm ul}=dT_0$ is equal to that of $\max\{t_{{\rm ul},k},k \in \mathcal{K}\}=dT_0$, which is given by
\begin{equation}
\Pr \big \{\max\{t_{{\rm ul},k},k \in \mathcal{K}\}=dT_0\big \}=\prod_{k=1}^K \Pr\{t_{{\rm ul},k}\leq dT_0\}- \prod_{k=1}^K \Pr\{t_{{\rm ul},k}\leq (d-1)T_0\}.   \label{eq22}
\vspace{1mm}
\end{equation}

Recalling \eqref{ap4}, $\Pr\{t_{{\rm ul},k}\leq dT_0\}=1-\Pr\{Z_d >0\}$. Hence, substituting \eqref{LRappa} into \eqref{eq22}, the distribution of $T_{\rm ul}$ can be expressed as
\begin{align}
&\Pr\{ T_{\rm ul}=dT_0 \}=\big(1-\Pr\{Z_{d}>0\}\big)^K-\big(1-\Pr\{Z_{d-1}>0\}\big)^K.\label{appblast}\nonumber \\
&=\bigg(\frac{1}{\sqrt{2\pi}}\!\int_{-\infty}^{\omega_d}e^{-\frac{u^2}{2}}du\!-\!\frac{e^{-\frac{\omega_d^2}{2}}}{\sqrt{2\pi}}\Big(\frac{1}{\psi_d}-\frac{1}{\omega_d}\Big)\bigg)^K \!\!-\bigg(\frac{1}{\sqrt{2\pi}}\! \int_{-\infty}^{\omega_{d-1}}\! e^{-\frac{u^2}{2}}du\!-\!\frac{e^{-\frac{\omega_{d-1}^2}{2}}}{\sqrt{2\pi}}\Big(\frac{1}{\psi_{d-1}}\!-\!\frac{1}{\omega_{d-1}}\Big)\bigg)^K.
\end{align}

\vspace{-5mm}

\section{Proof of Lemma 2}
Similar to the proof of Lemma 1, we can use saddle point approximation to get the distribution of $T_{\rm dl}^{(n)}$. Specifically, there is a difference between uplink and downlink rate distribution. Since $h_{\rm dl}=\min\{|h_{{\rm dl},k}|,k \in \mathcal{K}\}$, we can get the distribution of $h_{\rm dl}$, which is given by
\begin{align}
f_{h_{\rm dl}}(x)&= K\big[1-F_h(x)\big]^{K-1}f_h(x)\nonumber\\
&=Kxe^{-\frac{Kx^2}{2}}\mathbb{I}\{x>0\},\label{b1}
\end{align}
where $F_h(x)=\big(1-e^{\frac{-x^2}{2}}\big)\mathbb{I}\{x>0\}$ is the CDF of $|h_{{\rm dl},k}|$. Based on \eqref{b1}, the distribution of $\frac{r_{\rm dl}}{B_{\rm dl}}$ is given by
\begin{equation}
    f_{\rm dl}(r)=\frac{K}{2\lambda_{d}}e^{\frac{K}{2\lambda_{d}}}e^{r-\frac{K}{2\lambda_d}e^r}\mathbb{I}\{r \geq 0\},
\vspace{-2mm}
\end{equation}
where $\lambda_d=\frac{P_{\rm dl}}{\sigma^2}$ is the corresponding SNR. 

Then the MGF of $\frac{r_{\rm dl}}{B_{\rm dl}}$ is given by
\begin{equation}
    M_{\rm dl}(s)=e^{\frac{K}{2\lambda_d}}\Big(\frac{2\lambda_d}{K}\Big)^s\Gamma\Big(s+1,\frac{K}{2\lambda_d}\Big). \label{b3}
\vspace{-1mm}
\end{equation}
The CGF of $\tilde Z_d$ is given by
\begin{align}
    \tilde K_d(s)&=\frac{S}{B_{\rm dl}T_0}s+d\ln{M_{\rm dl}(-s)} \nonumber\\
    &=\frac{S}{B_{\rm dl}T_0}s+d\ln\left[e^{\frac{K}{2\lambda_d}}\Big(\frac{2\lambda_d}{K}\Big)^{-s}\Gamma\Big(-s+1,\frac{K}{2\lambda_d}\Big)\right].\label{b4}
\end{align}
Comparing \eqref{b4} with \eqref{ap9}, we can see that if we let $\tilde{\lambda}=\frac{\lambda_d}{K}$, then the further derivation is similar to the derivation of the uplink delay. Thus, we omit it due to the limitation of layout. Finally, the distribution of $T_{\rm dl}$ is given as shown in \eqref{lemma3.1}.

\section{Proof of Corollary 1}
Let $F(\cdot)$ denote the CDF of the user $k$'s one iteration delay $T_{k}^{(n)}$, which is given by 
\begin{equation}
F(x)
=\sum_{j=1}^{\lfloor \frac{x}{T_0} \rfloor}\sum_{i=1}^{j-1}\varrho(i)\upsilon(j-i).
\end{equation} 
Then the one iteration delay of the FL system can be expressed as $T^{(n)}=\max\big\{T_k^{(n)},k \in \mathcal{K}\big\}$. From \cite{extremevalue}, we can know that, if the limiting distribution of $T^{(n)}$ exists, it must belong to one of the three standard extreme value distributions: Fr{\'e}chet, Weibull and Gumbel distribution. The distribution of $T_{k}^{(n)}$ determines the limiting distribution of $T^{(n)}$. Therefore, according to Theorem 2.13 in \cite{extremevalue} and \cite{extreme2}, we can give the sufficient condition for the distribution of $T^{(n)}$ satisfying Gumbel distribution.

Let $w(F)$ denote the upper endpoint of $F(x)$, which is given by 
\begin{equation}
w(F)=\sup \{x:F(x)<1\}.
\end{equation}
Based on the definition of $T_{k}^{(n)}$, we can know that the $w(F)=\infty$. 
Hence, if there exist some finite $b$ satisfy
\begin{equation}
\int_{b}^{+\infty}\big(1-F(y)\big)dy < +\infty,
\vspace{-2mm}
\end{equation}
then for $0<t<+\infty$, we define
\vspace{1mm}
\begin{equation}
R(t)
=\frac{\int_{t}^{+\infty}\big(1-F(y)\big)dy}{1-F(t)} =T_0\Big\lceil \frac{t}{T_0} \Big \rceil-t+\frac{\sum_{d=\lceil \frac{t}{T_0} \rceil}^{\infty}\big(1-\sum_{j=1}^{d}\sum_{i=1}^{j-1}\varrho(i)\upsilon(j-i)\big)}{1-\sum_{j=1}^{\lfloor \frac{t}{T_0} \rfloor}\sum_{i=1}^{j-1}\varrho(i)\upsilon(j-i)},
\vspace{1mm}
\end{equation}
which is also known as mean residual life function. For all real $x$, if
\vspace{1mm}
\begin{equation}
\lim_{t \to +\infty}\frac{1-F\big(t+xR(t)\big)}{1-F(t)}=\lim_{t \to +\infty}\frac{1-\sum_{j=1}^{\lfloor \frac{t+xR(t)}{T_0} \rfloor}\sum_{i=1}^{j-1}\varrho(i)\upsilon(j-i)}{1-\sum_{j=1}^{\lfloor \frac{t}{T_0} \rfloor}\sum_{i=1}^{j-1}\varrho(i)\upsilon(j-i)}=e^{-x},
\vspace{1mm}
\end{equation}
then there are sequences $a_n$ and $b_n>0$ such that, $-\infty <x<\infty$,
\vspace{-1mm}
\begin{equation}
\lim_{K \to +\infty} \Pr \big \{T^{(n)}(K)<a_K+b_Kx\big\}=e^{-e^{-x}}. \label{prob}
\end{equation}
The constants $a_K$ and $b_K$ can be chosen as 
\begin{align}
a_K&=\inf \bigg \{x: 1-F(x)\leq \frac{1}{K}\bigg \},\\
b_K&=R(a_K).
\end{align}

\vspace{-3mm}
\section{Proof of Theorem 3}
To get the distribution of $T^{(n)}$ in the asynchronous downlink scheme, we need to get the one user's delay first, then we take the maximum order statistic of all user's delay. Similar to the synchronous downlink scheme, one user's uplink delay can be obtained by Lemma 1. Moreover, one user's downlink delay is given in Lemma 3. Thus, the delay of user $k$ in one iteration is given as the convolution of the uplink delay and downlink delay
\begin{align}
    \Pr\big\{ T_{k}^{(n)}=dT_0\big\}&=\sum_{i=1}^{d-1} \Pr\big\{ t_{{\rm ul},k}^{(n)}=iT_0\big\}\Pr\big\{ t_{{\rm dl},k}^{(n)}=(d-i)T_0\big\} \label{appd.1}\nonumber \\
&=\sum_{i=1}^{d-1}\varrho(i)\vartheta(d-i).
\end{align}
By substituting \eqref{eq20}, \eqref{zhat} into \eqref{appd.1}, the corresponding distribution is given by
\begin{align}
    \Pr\big\{ T^{(n)}=dT_0 \big\}&=\prod_{k=1}^K \Pr\big\{T_{k}^{(n)}\leq dT_0\big\}-\prod_{k=1}^K\Pr\big\{T_{k}^{(n)}\leq (d-1)T_0\big\}\nonumber\\
&=\bigg(\sum_{j=1}^{d} \sum_{i=1}^{j-1}\varrho(i)\vartheta(j-i)\bigg)^K-\bigg(\sum_{j=1}^{d-1} \sum_{i=1}^{j-1}\varrho(i)\vartheta(j-i)\bigg)^K.
\end{align}

\section{Proof of Corollary 2}
The most parts of the derivation are similar to Corollary 1. The difference lies in the CDF of $T_k^{(n)}$. In the asynchronous downlink scheme, the CDF of $T_k^{(n)}$ is given by
\begin{equation}
F(x)=\sum_{j=1}^{\lfloor \frac{x}{T_0} \rfloor} \sum_{i=1}^{j-1}\varrho(i)\vartheta(j-i).
\end{equation}
Then we can get a similar result as derived from Corollary 1.

\section{Proof of Theorem 4}
Our proof relies on the LDT. Having the distribution of the one iteration delay $T^{(n)}$, we can analyze the tail distribution of $T_c$ by LDT.  
Based on Theorem 23.3 in \cite{LDT}, first we calculate the CGF of $T^{(n)}$ as
\begin{align}
K(s)=\ln \mathbb{E}\big\{e^{sT^{(n)}}\big\}=\ln \bigg( \sum_{d=1}^{\infty} e^{sdT_0} \sum_{i=1}^{d-1} \varphi(i)\upsilon(d-i)\bigg).
\end{align}
Then the rate function $I(s)$, which is the Legendre transform of $K(s)$, is given by 
\begin{align}
\Lambda^*(x)=\underset{s \in \mathbb{R}}{\sup} \big (sx-K(s) \big )
=s^\star x-\ln \bigg( \sum_{d=1}^{\infty} e^{s^\star dT_0} \sum_{i=1}^{d-1} \varphi(i)\upsilon(d-i)\bigg),
\end{align} 
where $s^\star$ can be calculated as
\begin{equation}
\frac{\sum_{d=1}^{\infty} dT_0 e^{s^\star dT_0} \sum_{i=1}^{d-1} \varphi(i)\upsilon(d-i)}{\sum_{d=1}^{\infty} e^{s^\star dT_0} \sum_{i=1}^{d-1} \varphi(i)\upsilon(d-i)}=x.
\end{equation}
Then, for every $x> \mathbb{E}\{T^{(n)}\}$,
\begin{equation}
\lim_{n \to \infty} \frac{1}{n}\ln \Pr\bigg\{ \sum_{i=1}^{n} T^{(i)} \geq xn\bigg\}=-\Lambda^*(x).
\end{equation}
Hence, for $x> \mathbb{E}\{T^{(n)}\}$ and $n \to \infty$, the tail distribution of $T_c$ can be expressed as
\begin{equation}
\Pr\{ T_c \geq xn\}=e^{-n\Big(s^\star x-\ln \big( \sum_{d=1}^{\infty} e^{s^\star dT_0} \sum_{i=1}^{d-1} \varphi(i)\upsilon(d-i)\big)\Big)}.
\end{equation}
Recalling that $I_0$ is the upper bound of $n$, it can be seen as a relatively large number. Hence, for $\tau \geq I_0 \mathbb{E}\{T^{(n)}\}$, by inserting $n=I_0$ and $\tau=xI_0$, \eqref{th4.1} holds.

\section{Proof of Theorem 6}
According to \eqref{co1} and \eqref{co2}, the probability density function of $T^{(n)}$ when $K$ approaches infinity is given by
\begin{equation}
f(y)=\frac{1}{b}e^{-\frac{y-a}{b}-e^{-\frac{y-a}{b}}}.
\end{equation}
Then similar to the derivation in Theorem 4, we can get
\begin{equation}
K(s)=-\ln b+\frac{a}{b}-e^{\frac{a}{b}} +\ln \bigg( \int_{0}^{+\infty} e^{(s-\frac{1}{b})y } e^{-e^{-\frac{1}{b}y}}dy \bigg),
\end{equation}
\begin{equation}
\Lambda^*(x)=\underset{s \in \mathbb{R}}{\sup} \big (sx-K(s) \big )
=s^\star x-\ln \bigg( \int_{0}^{+\infty} e^{(s^\star-\frac{1}{b})y }e^{-e^{-\frac{1}{b}y}}dy \bigg)+\ln b-\frac{a}{b}+e^{\frac{a}{b}},
\end{equation}
where $s^\star$ satisfies
\begin{equation}
\int_{0}^{+\infty} ye^{(s^\star-\frac{1}{b})y }e^{-e^{-\frac{1}{b}y}}dy=x\int_{0}^{+\infty} e^{(s^\star-\frac{1}{b})y }e^{-e^{-\frac{1}{b}y}}dy.
\end{equation}
Hence, for $\tau \geq I_0\mathbb{E}\{T^{(n)}\}$, the tail distribution of $T_c$ can be expressed as
\vspace{1mm}
\begin{align}
\Pr\{ T_c \geq \tau \}&=\exp\Bigg(-I_0\bigg(s^\star x-\ln \Big( \int_{0}^{+\infty} e^{(s^\star-\frac{1}{b}-e^{-\frac{1}{b}})y }dy\Big)+\ln b-\frac{a}{b}+e^{\frac{a}{b}}\bigg)\Bigg)\nonumber\\
&=\exp(- s^\star \tau)\bigg(\frac{e^{\frac{a}{b}-e^{\frac{a}{b}}}}{b}\int_{0}^{+\infty} e^{(s^\star-\frac{1}{b})y }e^{-e^{-\frac{1}{b}y}} dy\bigg)^{I_0}.
\end{align}

\vspace{-3mm}
\section{Proof of Corollary 4}
Once we get the distribution of the number of iterations for convergence $N$, we obtain its PGF $G_N(z)$. From Theorem 2 and Theorem 3, we can also get the PGF of one iteration delay, which we denote as $G_I(z)$. Therefore, the MGF of overall delay is given by
\begin{align}
G_c(z)&=\sum_{d=1}^{\infty} z^{dT_0}\Pr\{T_c=dT_0\} \nonumber \\
&=\sum_{j=1}^{\infty} \Pr\{N=j\}  \sum_{d=1}^{\infty} \Pr\{T_c=dT_0|N=j\}z^{dT_0} \nonumber \\
&=\sum_{j=1}^{\infty} \Pr\{N=j\} \big(G_I(z)\big)^j \nonumber \\
&=G_N\big( G_I(z)\big).
\end{align}

\vspace{-1mm}
Based on the $G_c(z)$, we can derive the distribution of $T_c$ from 
\begin{equation}
\Pr\{T_c=dT_0\}=\frac{G^{(d)}_c(0)}{d!}=\frac{G^{(d)}_n\big( G_I(z)\big)\big|_{z=0}}{d!}.
\vspace{-2mm}
\end{equation}



\bibliographystyle{IEEEtran}
\begin{spacing}{0.84}
\bibliography{mybib}
\end{spacing}

\ifCLASSOPTIONcaptionsoff
  \newpage
\fi

\end{document}